\documentclass[superscriptaddress,twocolumn,aps,prx,10pt]{revtex4-1}

\usepackage{times}
\usepackage{amsmath}
\usepackage{amsfonts}
\usepackage{amssymb}
\usepackage{bm}
\usepackage{graphicx}
\usepackage{color}
\usepackage{float}
\usepackage{enumitem}   

\begin{document}

\title{Active glasses}

\author{Liesbeth M.~C.~Janssen}
\email[Electronic mail: ]{L.M.C.Janssen@tue.nl}
\affiliation{Theory of Polymers and Soft Matter, Department of Applied Physics, Eindhoven University of Technology,
             P.O. Box 513, 5600MB Eindhoven, The Netherlands}

\date{\today}

\begin{abstract}
Active glassy matter has recently emerged as a novel class of non-equilibrium soft 
matter, combining energy-driven, active particle movement with dense and
disordered glass-like behavior. Here we review the state-of-the-art in this
field from an experimental, numerical, and theoretical perspective. 
We consider both non-living and living active glassy systems, and discuss how several hallmarks of glassy dynamics 
(dynamical slowdown, fragility, dynamical
heterogeneity, violation of the Stokes-Einstein relation, and aging) are manifested in such materials.
We start by reviewing the recent experimental evidence in this area of research, followed by an overview of the main numerical simulation studies 
and physical theories of active glassy matter. We conclude by outlining several open questions and possible directions for future work.
\end{abstract}

\maketitle

\section{Introduction}

\subsection{Active and glassy matter}
During the last decade, active matter has emerged as a new and rapidly
expanding research area within the field of condensed matter science
\cite{Marchetti2013, Ramaswamy2010, Bechinger2016}. The term
'active matter' refers to materials whose constituent particles (or 'agents')
are capable of converting energy into some form of autonomous motion. In
general, the energy can either be stored within the particles themselves or
supplied externally, e.g. by introducing a chemical fuel or an external
electromagnetic field. The type of active particle motion may occur in the
translational, vibrational, or rotational degrees of freedom, or a combination
thereof. Examples of active matter in the natural world are abundant, ranging
from macroscopic organisms such as flying birds and swimming fish to the
microscopic realm of motile bacteria and cells, down to the subcellular level
of e.g.\ the cytoskeleton and molecular motor proteins. In all these cases, ATP
is the main fuel source. From the synthetic side, active systems are now
also available across many length scales; examples of such man-made structures
include electrically-driven robots, granular particles on a vibrating table \cite{Narayan2007},
catalytic \cite{Howse2007} and light-activated \cite{Palacci2013} colloids, metal-capped colloids in
near-critical mixtures \cite{Buttinoni2013}, swimming oil droplets \cite{Maass2016}, catalytic stomatocyte
nanoparticles \cite{Wilson2012}, and artificial molecular motors \cite{Koumura1999, Kelly1999}.
Importantly, since each agent in an active-matter system is constantly
consuming energy to generate its own movement, the material is said to be out
of thermodynamic equilibrium \textit{at the single-particle level}. This is to
be contrasted with many other methods to bring a material out of equilibrium,
such as a sudden quench of a thermodynamic control parameter (e.g.\ temperature
or density) or the application of an external force field (e.g.\ shear); 
these latter protocols 
do not act on the scale of individual particles, but rather on the material as a whole 
or on the boundaries.
 
One of the central goals in active-matter physics research is to explore and
understand how the intrinsic non-equilibrium nature of active particles can
give rise to complex, collective, and novel self-organizing behavior that is
absent in the passive counterpart. Let us first briefly consider the case of a
single, \textit{non-interacting} active Brownian particle. Both theory and experiment show
that such an active colloid will undergo normal diffusion
just as a conventional 'passive' Brownian particle, at least at time scales comparable to (or larger than) the particle's
rotational diffusion time \cite{Howse2007}. In this diffusive regime, the only
effect of the self-propelled motion is that the effective diffusion constant
will be larger than in the passive case (with an enhancement term
proportional to the self-propulsion speed squared \cite{Howse2007})--a result that is often
interpreted as an effectively higher temperature $T_{eff}$. Thus, in this example,
an active material appears to be rather similar to an effective equilibrium system.

The physics can
change profoundly, however, when active particles become governed by mutual
(two- or many-body) interactions. Since the pioneering 1995 study by Vicsek and
co-workers--which provides a minimal model for flocking behavior due to
aligning interactions \cite{Vicsek1995}--, numerous novel phenomena in
interacting active-matter systems have been discovered, including
Motility-Induced Phase Separation \cite{Redner2013,Cates2015}, active
turbulence \cite{Wensink2012, Rabani2013, Giomi2015, Wu2017}, active nematic
liquid-crystalline behavior \cite{Keber2014,Doostmohammadi2018}, spontaneous
compartmentalization \cite{Spellings2015,Grauer2018}, and synthetic quorum
sensing \cite{Bauerle2018}. These findings not only call for new developments 
in fundamental non-equilibrium condensed matter physics, but they also offer 
new possibilities in biology and materials science. Indeed, as will be discussed later in this
review, the framework of active matter provides a unique angle of
approach to describe the complex phenomenology of living systems
from a novel statistical-physics-based point of view. Furthermore, the incorporation of activity in
synthetic systems offers unprecedented possibilities to create functional materials with 'smart' and
life-like properties that would be unattainable in thermodynamic
equilibrium. Overall, the field of active matter thus offers an exciting new paradigm
with relevance in both the fundamental and applied sciences 
\cite{Marchetti2013, Ramaswamy2010, Bechinger2016, Trepat2018, Needleman2017}.

\begin{figure*}
        \begin{center}
    \includegraphics[width=0.95\textwidth]{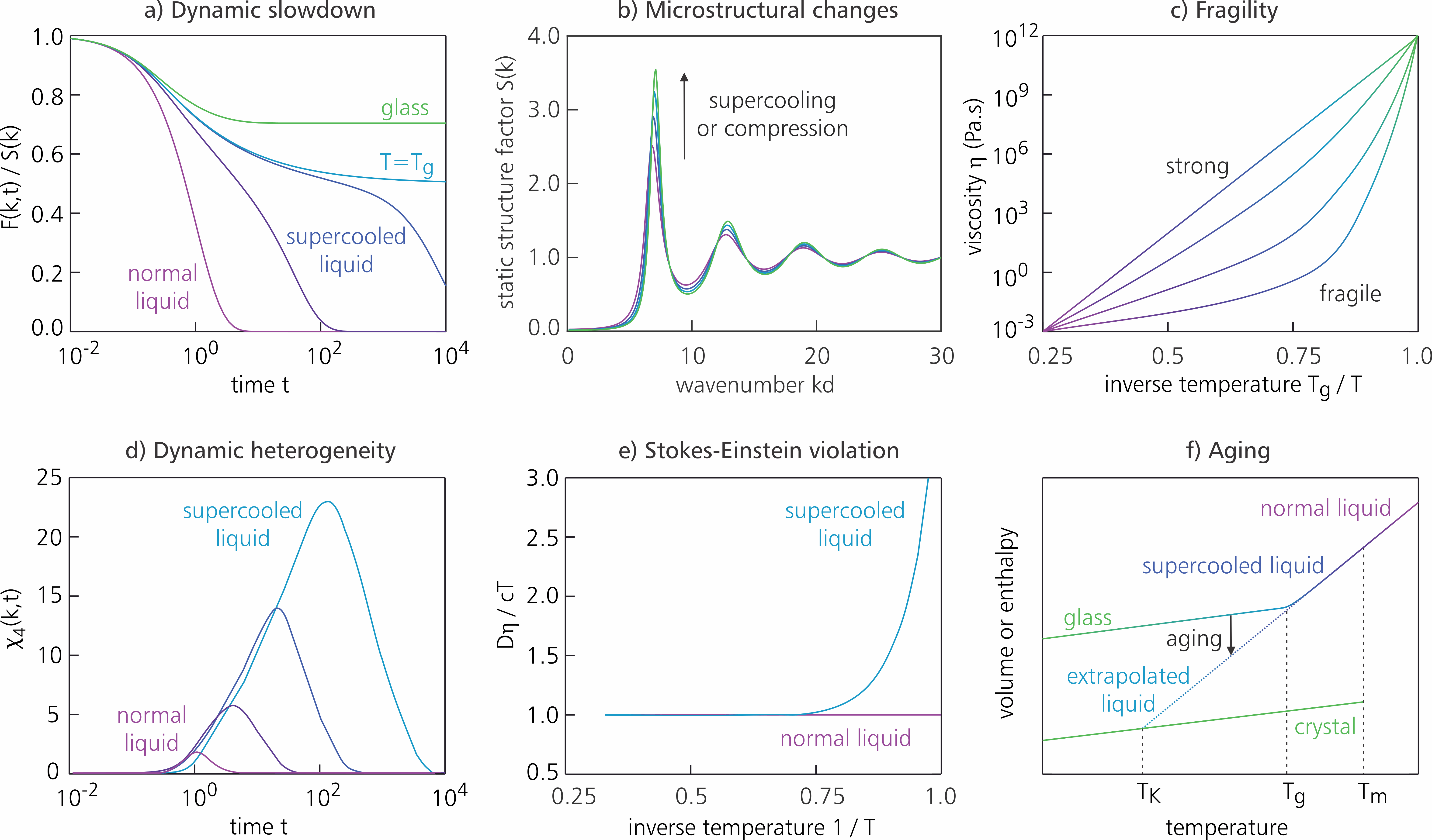}
  \end{center}
  \caption{
  \label{fig:hallmarks}
  Illustration of the main hallmarks of glassy dynamics. (a) Dramatic dynamical
slowdown upon supercooling or compression, as quantified by the collective
intermediate scattering function $F(k,t)$. This function probes correlations
among density modes at a certain wavenumber $k$ and over a time $t$; its
characteristic relaxation time is also a measure for the viscosity.  At the
glass transition temperature $T_g$, $F(k,t)$ fails to decay on any practical
time scale, thus signaling the formation of a solid state. Note the logarithmic
time scale.  (b) Microstructure of a glass-forming material, as quantified by
the static structure factor $S(k)$. Here $S(k)$ is obtained from the
Percus-Yevick approximation for hard spheres; $k$ is given in units of the
inverse particle diameter $d$.  Only subtle structural changes appear during
glass formation, most notably in the main peak of $S(k)$. The growth of this
peak is associated with the cage effect.  (c) Angell plot or fragility plot
showing the increase of the viscosity as a function of normalized inverse
temperature. Strong glass formers exhibit an Arrhenius-type growth, while
fragile glass formers solidify more abruptly in a super-Arrhenius fashion. Note
the logarithmic viscosity scale.  (d) Dynamical heterogeneity, as quantified by
the dynamical susceptibility $\chi_4(k,t)$. This function essentially probes
\textit{fluctuations} in $F(k,t)$; the size of the peak of $\chi_4(k,t)$ is a
measure for the number of cooperatively rearranging particles.  (e) Breakdown
of the Stokes-Einstein relation $D\eta/cT$, where $c$ is a constant. In the
supercooled regime, the diffusion constant $D$ and viscosity $\eta$ become
decoupled.  (f) Aging dynamics in the glassy state. After a glass has formed at
temperature $T_g$, the material may slowly evolve toward the extrapolated
equilibrium supercooled liquid branch below $T_g$. This branch starts at the
melting temperature $T_m$ and terminates, presumably, as the Kauzmann
temperature $T_K$.
}
\end{figure*}

A different branch of condensed matter physics concerns the study of glasses \cite{Binder2011, Berthier2011a,Royall2015, Berthier2016}.
Glassy materials exhibit solid-like behavior but, unlike crystalline solids,
they lack any long-range structural order. The study of such disordered solids
has a long history, with the first man-made glasses dating back to ca.\ 3500
BC \cite{Beck1934,Moorey1999}. It is now widely accepted that any material can, in principle, exist in a
glassy state; indeed, our modern society makes use of a wide variety of
amorphous solids, including organic, inorganic, polymeric, metallic, and
colloidal glasses. The most common method of producing a glass is to supercool
or compress a liquid until the viscosity $\eta$ (or structural relaxation time
$\tau$) exceeds a certain threshold value; if crystallization is avoided, the
resulting material can then be regarded as an amorphous solid on any practical
time scale. The temperature or density at which the viscosity reaches the
solidification threshold (typically defined as $10^{12}$ Pa.s) is known as the
glass transition. At this transition, the supercooled liquid is said to have
fallen out of equilibrium into a non-ergodic glassy state. Curiously, while the
experimental process of glass formation has been known for centuries, it is
governed by a multitude of complex phenomena that remain notoriously poorly
understood to this day \cite{Berthier2011a}. In fact, the nature of the glassy state and the glass
transition has been called "the deepest and most interesting unsolved problem
in solid state theory" \cite{Anderson1995}, and in 2005 the journal Science declared it one of the
"most compelling puzzles and questions facing scientists today" \cite{Science2005}. There are
several excellent reviews which detail the experimental phenomenology and
current theoretical understanding of glassy materials \cite{Berthier2011a,Royall2015,Tarjus2011, Biroli2013, Langer2014}; for the purpose of this
paper, we briefly summarize the main hallmarks of vitrification below.

\subsection{Hallmarks of glassy dynamics}
\label{sec:hallmarks}

Among the many complex phenomena associated with glass formation, we address
five of them in this review: i) dramatic dynamical slowdown, ii) fragility, iii)
dynamical heterogeneity, iv) violation of the Stokes-Einstein relation, and v)
aging (see Fig.\ \ref{fig:hallmarks}). The first aspect is arguably the most striking hallmark of
vitrification: as a liquid is supercooled toward the glass transition, its
viscosity or relaxation increases by many orders of magnitude upon only a mild
decrease in temperature. At the same time, however, this spectacular dynamical
slowdown is accompanied by only subtle changes in the microstructure of the
material \cite{Royall2015}. Indeed, the structure of a supercooled liquid or glass is almost
indistinguishable from that of an ordinary liquid (as quantified by e.g.\ the
static structure factor \cite{Hansen2013}). It is this apparent disconnect between structural and
dynamical properties that lies at the heart of the glass transition problem:
there is still no theory to accurately and rationally link the microstructure
of a supercooled liquid to its quantitative relaxation dynamics. At a
\textit{qualitative} level, however, the dynamical slowdown has been successfully
explained by theories such as Mode-Coupling Theory \cite{Gotze2008} in terms of the so-called cage
effect \cite{Kob2002, Weeks2002}. This effect signifies that, as the density increases or temperature
decreases, particles become trapped in transient cages formed by their
neighboring particles. The hindered particle motion of a
caged particle in turn also facilitates the effective caging of its
neighbors, culminating into a highly non-linear dynamical slowdown upon only a
small change in density or temperature. 

The second aspect, fragility, refers to the fact that not all materials vitrify
in the same manner \cite{Angell1995,Tarjus2014}. So-called 'strong' glass formers solidify rather gradually,
exhibiting an Arrhenius-type growth of the viscosity upon supercooling, while
'fragile' materials vitrify more abruptly in a super-Arrhenius fashion. Many
materials fall in between these two extremes, and in fact there are numerous
examples of systems that also exhibit a fragile-to-strong crossover \cite{Elmatad2011}. Although
there is consensus that network-forming materials, such as silica, tend to
behave as strong glass formers, and that materials dominated by isotropic
particle interactions, such as colloidal hard spheres, are generally more
fragile \cite{Ozawa2016}, a microscopic framework to predict the fragility for a
given material composition and microstructure is still lacking \cite{Kelton2017}. It is widely
believed that a resolution to this problem, i.e.\ obtaining a detailed
understanding of the microstructural origins of fragility, will be key in
ultimately achieving a universal description of the glass transition. 

The third aspect, dynamical heterogeneity, is a fairly recent addition to the
phenomenology of glassy dynamics \cite{Ediger2000,Berthier2011,Berthier2011c}. Dynamical heterogeneity signifies that the
structural relaxation dynamics of a supercooled liquid does not proceed
uniformly across the entire material, but rather in groups of collectively
rearranging particles while the rest of the system remains temporarily frozen.
The appearance of such mobile clusters fluctuates both in space and in time (at
a fixed supercooled temperature and density), and the cluster size tends to
grow upon approaching the glass transition. In a qualitative sense, this can be
understood as a consequence of caging, requiring an increasingly large
collective effort to mobilize particles as the density increases. The notion of
cooperatively rearranging domains dates back to the 1965 work of Adam and
Gibbs \cite{Adam1965}, but it was not until the late 1990s that dynamical heterogeneity was
firmly established in simulation \cite{Hurley1995,Kob1997a} and experiment \cite{Kegel2000,Weeks2000}. While it is now widely
believed that the size of dynamically heterogeneous regions represents an
important dynamical length scale in the vitrification process, it is not yet
established whether this length scale will ultimately diverge at a finite
temperature to signal a critical phenomenon \cite{Berthier2005,Albert2016}.

A somewhat related aspect of glassy dynamics is violation of the
Stokes-Einstein relation \cite{Hodgdon1993,Tarjus1995,Shi2013,Charbonneau2014,Charbonneau2018}. This equation states that the viscosity $\eta$ (or
relaxation time $\tau$) and diffusion constant $D$ are related to the
temperature $T$ as $\eta D/T = constant$. In ordinary equilibrium liquids, the
Stokes-Einstein relation is generally obeyed; in most supercooled liquids,
however, the viscosity increase is stronger than the diffusion-constant
decrease, and such Stokes-Einstein violation becomes more pronounced as the
glass transition temperature is approached. The intuitive explanation for this
decoupling is that diffusion is governed by the fastest particles, whereas structural
relaxation is dominated by the slowest ones \cite{Ediger2000}. Note that the simultaneous 
existence of both fast and slow particle populations is also the key aspect of dynamically heterogeneous behavior, 
and hence it is generally assumed that Stokes-Einstein violation in glassy liquids
is essentially a manifestation of dynamical heterogeneity \cite{Zangi2007,Sengupta2013}.
While it has been suggested that such phenomenology may arise from critical dynamical fluctuations \cite{Biroli2007}, 
it has also been proposed that non-critical hopping processes, i.e.\ the effective escape of particles from their
local cages, predominantly underlie the observed viscosity-diffusion decoupling \cite{Charbonneau2014}.  
Overall, a general first-principles framework to accurately predict both the degree of Stokes-Einstein violation
and dynamical heterogeneity for a glass-forming material is still missing.

\begin{figure}
        \begin{center}
    \includegraphics[width=0.48\textwidth]{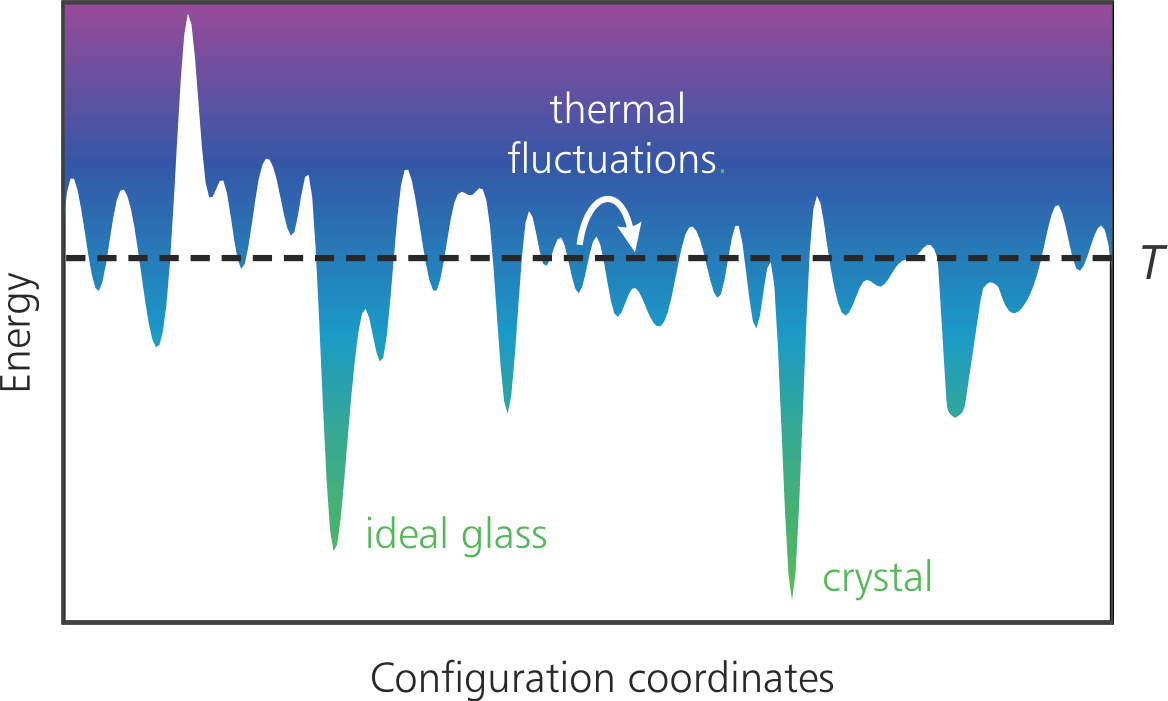}
  \end{center}
  \caption{
  \label{fig:energy}
  Illustration of the energy landscape picture. 
 The $x$-axis represents all configurational coordinates of an $N$-particle system. 
 The global minimum of the energy is assumed to be the crystalline state, while
the lowest possible energy state for a disordered configuration is the ideal
glass. Note that this example assumes that only a single crystalline state exists. 
The dashed line indicates a typical temperature $T$ at which thermal
fluctuations may allow the system to surmount local barriers.  The figure is
adapted from Ref.\ \cite{Janssen2017}.
  }
\end{figure}

The fifth aspect, aging, refers to the fact that the behavior of a material can
exhibit an explicit dependence on its age (at a fixed temperature and density). That
is, the structural and dynamical properties may slowly evolve as the material
becomes older \cite{Struik1977,Berthier2009,Micoulaut2016}. Aging is generically understood as a non-equilibrium phenomenon
which signals a material's gradual approach toward an underlying equilibrium
state. A convenient paradigm to describe this behavior is the so-called energy-landscape picture \cite{Stillinger1995,Debenedetti2001,Heuer2008,Raza2015}, 
which represents the system's total (free) energy as a highly rugged surface in a 
high-dimensional configuration space (see Fig.\ \ref{fig:energy}). The process of aging then corresponds to 
the system's progressive exploration of deeper energy minima on this surface \cite{Kob1997,Kob2000,Rehwald2010}. 
In supercooled liquids, aging effects are typically only observed after
a (small) temperature quench; if the liquid is supercooled sufficiently slowly,
it will behave as an ordinary equilibrium liquid in the sense that, e.g.,
ergodicity and the fluctuation-dissipation theorem hold. In the glass state,
however, the relaxation time that is needed to reach (quasi-)equilibrium will
exceed--by definition--any practical time scale, implying that the supercooled
liquid has fallen out of equilibrium. Physical aging is therefore typically
observed within the glassy phase, and is a manifestation of the material's
tendency to reach a lower energy state \cite{Struik1977}. Ultimately, this aging
behavior will bring the material into a deeply supercooled (quasi-equilibrium)
liquid phase at a temperature below the original glass transition temperature.
It remains to be established whether the equilibrium supercooled branch will
eventually terminate at a low but finite temperature (the so-called Kauzmann
temperature $T_K$ \cite{Kauzmann1948}); this point would then correspond to the lowest theoretically
possible glass transition temperature--a temperature at which the
configurational entropy of the glass should rigorously vanish. 
The difficulty in testing
this hypothesis is that it would require, in principle, infinitely slow
supercooling rates and/or exceptionally long aging and equilibration times
below the operational glass transition temperature \cite{Royall2018}; such time scales inherently
exceed the time scale of any practical simulation or experiment. For a very recent and comprehensive discussion
on the configurational entropy of glass-forming liquids, we refer the reader to Ref.\ \cite{Berthier2019a}.

\subsection{Active glassy matter}

On top of the already rich physics of both glassy and active matter, recent
years have witnessed the emergence of a new subfield that lies at the interface
of these two classes of materials: active glassy matter. Such materials are
comprised of energy-driven motile particles that collectively exhibit dense and
disordered glass-like behavior. Hence, active glassy matter combines multiple
distinct non-equilibrium properties in a single material, offering an exciting
playground for the discovery of fundamentally new physics. In fact, there is
now a growing realization that active glassy physics is manifested in many
biological systems, and that glassy behavior in living cells may even carry a
biological function. In the context of materials science, the combination of
activity and glassiness may provide new possibilities to create 'smart'
amorphous materials with life-like and adaptive functionalities. For example,
considering that a particle's self-propelled motion can be activated or de-activated by          
external cues, an active glassy material may be (locally) fluidized and re-solidified   
in an externally controllable manner. This in turn holds application potential for, e.g., 
shape-shifting and phase-changing materials, soft robotics, switchable sensors, self-healing glasses,  
and on-demand storage and release functionalities. 
Although the
experimental realization of such synthetic active glasses is still in its
infancy, the theoretical study of active glassy matter has already seen
exciting progress in recent years. This review aims to provide an overview of
this emergent field, highlighting the experimental evidence and current
theoretical understanding of active glass-like behavior in both artificial and
biological systems. We will pay special attention to the manifestation of the
five main hallmarks of glassy dynamics in non-equilibrium active matter, and
discuss whether the underlying physics of glassy phenomenology may be similar
or distinct in passive and active materials.

Before we end this section, let us make a general remark on the distinction
between glassy and jamming physics. Both phenomena describe a transition
between a fluid-like and solid-like state, and the terminology is often used
interchangeably to describe the emergence of rigidity in disordered materials.
Here we use the convention  \cite{Parisi2010,Berthier2019} that the glass transition is a
\textit{dynamical} transition, corresponding to full kinetic arrest and loss
of ergodicity. Broadly speaking, this phenomenon arises from a competition
between particle crowding (due to e.g.\ increased density and caging) and the
particles' ability to move (due to e.g.\ thermal fluctuations, Brownian motion,
or self-propelled active motion). Jamming, on the other hand, is interpreted 
as a \textit{geometric} transition that is governed by an increase in the particle 
connectivity or the number of direct particle contacts \cite{VanHecke2009}. 
More generally, the jamming transition occurs in the
absence of any kind of (thermal, Brownian, or active) dynamics, and is defined as a 
zero-temperature and zero-activity
limit of the glass transition. Indeed it was recently argued by Berthier,
Flenner, and Szamel that strictly speaking, therefore, active particles can
undergo jamming only when they are not active \cite{Berthier2019}. Throughout this paper,
we will thus refer to aspects such as the dynamical slowdown as \textit{glassy},
rather than jamming, phenomena. 

This paper is organized as follows. In Sec.\ \ref{sec:expt}, we first review the experimental evidence
for glassy behavior in active matter. We start with very recent experiments on synthetic systems, followed by a
description of the observed intra- and intercellular glassy dynamics in living cells, respectively.
Section \ref{sec:sim} focuses on 'numerical experiments', i.e., computer simulation studies of active glassy matter.
Here again we first consider non-living systems and subsequently turn to the modeling of living cell tissues.
Section \ref{sec:theory} is devoted to recently developed theories of active glassy matter, including active versions
of spin-glassy theory, Mode-Coupling Theory, and Random First Order Transition Theory. In all cases, we focus mainly 
on the manifestation of the five hallmarks of glassy dynamics (Sec.\ \ref{sec:hallmarks}) in active materials. 
We summarize our key findings
in Sec.\ \ref{sec:concl} and conclude with an overview of open questions and possible directions for future research.

\section{Experimental studies of active glassy matter}
\label{sec:expt} 

\subsection{Synthetic active glasses}
\label{sec:exptsynth}
We begin by briefly reviewing the experimental work on artificial active glassy matter.
From the synthetic materials side, it has thus far proven challenging to
achieve self-propelled particles at sufficiently high densities such that
glassy dynamics can be observed, and hence the number of experimental studies
is still very limited.  Indeed, the first experiment on active colloidal
glasses was presented only very recently by Klongvessa \textit{et al}.\ \cite{Klongvessa2019,Klongvessa2019a} in a
sedimentation experiment of peroxide-driven colloids. By analyzing the dynamics
in a two-dimensional layer as a function of density and activity (measured in
terms of an effective temperature $T_{eff}$), it was found that increasing the
density always leads to slower structural relaxation, i.e.\ more glassy behavior. Interestingly, however, they also
observed a non-monotonic dependence of the relaxation time on $T_{eff}$: when
the passive system becomes \textit{weakly} active, the dynamics slows down, but
as the activity further increases the dynamics speeds up. This non-monotonic
effect was only observed in the high-density regime, i.e.\ at densities where
the passive material behaves as a glass. Thus, it appears that the role of
activity in dense amorphous colloidal systems cannot be mapped onto an
effectively passive system in a simple manner. 

In addition to dense active colloids, we also mention another promising class
of synthetic non-equilibrium materials that may offer an experimental
realization of active glassy matter, namely driven granulates. Such systems can
be realized experimentally by placing granular particles on, e.g., an
air-fluidized or vibrating bed. It is hoped that future experimental research
in this direction will shed more light on the non-trivial effects of
non-equilibrium self-propulsion in model glass-forming systems. In particular,
such studies should allow for better physical insight that may be less
straightforward to achieve in more complex systems such as living cells, and
may ultimately pave the way toward the development of synthetic materials with
new adaptive functionalities. 

\subsection{Living active glasses}
\label{sec:exptlife}

\subsubsection{Intracellular dynamics}
Let us now discuss the experimental evidence for active glassy behavior in
living systems. One of the oldest demonstrations of active-glass formation is
cryopreservation of cells, which amounts to the rapid freezing of living cell
tissue at cryogenic temperatures \cite{Pegg2007}. During this process, water inside and outside
the cells vitrifies into amorphous ice to effectively solidify the entire
tissue, thereby "stopping biological time" and rendering the material suitable
for long-term storage. The active components of the cell, such as the
cytoskeleton, thus become arrested within a water-rich glassy matrix. It is
perhaps interesting to note that, while the apparent lack of microstructural changes
upon supercooling makes glass formation one of the most notorious problems in
condensed matter physics, it is precisely this aspect that allows cells to
conveniently preserve their delicate internal structure as they undergo
solidification. Importantly, as in the case of ordinary glass formation, it is
crucial that crystallization is avoided during the supercooling process. For
living cells, the formation of ice crystals will in fact cause irreparable
damage to the tissue, and hence so-called cryoprotectants are commonly added to
inhibit water crystallization during cryogenic cell treatment \cite{Akiyama2019}.  

The explicit link between intracellular cytoplasmic properties and the main
hallmarks of glassy physics--including the dynamical slowdown and
fragility--was first recognized in 2009 by Fredberg, Weitz, and
co-workers \cite{Zhou2009}. They found that eukaryotic cells under osmotic stress exhibit an
orders-of-magnitude increase of the cytoplasmic viscosity with increasing
density. The observed dynamical slowdown conforms to an Arrhenius-type behavior,
making the cell's fragility reminiscent of strong glass formers such as silica
or soft deformable spheres \cite{Mattsson2009}. It was also discovered, however, that
ATP depletion significantly modulates the glass transition behavior of the
cell, indicating that non-equilibrium activity plays a non-trivial role in the
cellular glassy dynamics.
 
A 2013 study by Parry \textit{et al}.\ \cite{Parry2014} showed that the bacterial cytoplasm
also exhibits numerous glass-like properties. Unlike eukaryotic cells, the
bacterial cell interior lacks cytoskeletal motor proteins, and the transport
properties within the bacterial cytoplasm are therefore thought to be governed
primarily by crowding effects. Parry \textit{et al}.\ found that the cytoplasm
of metabolically inactive bacteria is glass-like in the sense that the
transport dynamics becomes extremely slow and dynamically heterogeneous. This
constitutes arguably the first demonstration of glassy dynamical heterogeneity at
the intracellular level. Interestingly, the study also revealed that metabolic
activity fluidizes the cytoplasm, implying that non-equilibrium active
processes offer a means to \textit{control} the bacteria's glassiness. Indeed, it was
hypothesized that the cell's ability to reversibly switch from a fluid-like to
more dormant, glass-like state may enable bacteria to survive in a
nutrient-poor environment. 

The 2017 work by Nishizawa \textit{et al}.\ \cite{Nishizawa2017} provided more insight into the
fragility of the intracellular cytoplasm. They studied both eukaryotic and prokaryotic
cells, and compared the glassy dynamics of both living samples and \textit{in vitro} models
from which the metabolic components and cytoskeletons were removed. Interestingly, 
it was found that the viscosity increase of the \textit{in vitro} cytoplasms upon crowding
conforms to a fragile, super-Arrhenius pattern; metabolically active living cells,
on the other hand, exhibited a strong, Arrhenius-type growth of the viscosity.
This confirms that activity can lead to a qualitatively different behavior of transport
properties within cells; in particular, the intracellular fragility of the living cell
appears to be fundamentally distinct from its non-living counterpart. A recent simulation study
by Oyama \textit{et al}.\ \cite{Oyama2019} rationalized the experimental observations of
Parry \textit{et al}.\ \cite{Parry2014} and Nishizawa \textit{et al}.\ \cite{Nishizawa2017}
in terms of an ATP-driven conformational change of proteins within the cytoplasm, thus 
offering a minimal model in which activity leads to volume fluctuations of particles, 
rather than self-propelled motion. These simulations revealed that only a small change in protein volume is 
sufficient to fluidize the glassy state and affect the cytoplasmic fragility. 

Recent work suggests that more complex organisms also employ glass-like
behavior as a biological self-protection mechanism. A 2017 study by Boothby
\textit{et al}.\ \cite{Boothby2017} found that tardigrades--multicellular micro-organisms which
can survive under extremely harsh conditions--undergo intracellular
vitrification as they are confronted with dehydration. The process relies on
the vitrification of intrinsically disordered proteins within the tardigrade
cytoplasm; the resulting glassy mixture may subsequently act as a protective
matrix for other dehydration-sensitive cellular components. Since the
vitrification process is reversible by rehydration, it is plausible that this
liquid-glass-like transition indeed acts as an effective survival strategy to
support the extreme resilience of tardigrades.

\subsubsection{Intercellular dynamics}
There is now a growing body of literature which suggests that glassy physics is
also manifested at the \textit{intercellular} level in e.g.\ amorphous
confluent cell sheets (i.e.\ cell layers with a packing fraction of unity). One of the first quantitative studies on glassy
collective cell dynamics is the 2011 work by Angelini \textit{et al}.\ \cite{Angelini2011}. They
demonstrated that the relaxation dynamics of two-dimensional epithelial
Madin-Darby Canine Kidney (MDCK) layers slows down as the cell density
increases, and that the cells' self-diffusion coefficient grows in a
non-Arrhenius fashion with increasing density. The corresponding fragility
index is comparable to that of a moderately fragile glass former. It was also
found that the glassy cell dynamics is manifestly heterogeneous, and that the
estimated correlation length associated with this dynamical heterogeneity
increases with density. Finally, since the density increase essentially
corresponds to a maturation of the cell layer, the observed glassy behavior may
be interpreted as a sign of aging. While at a phenomenological level these
features show a clear resemblance with ordinary glassy materials, the
relaxation dynamics in confluent cell layers has, of course, a fundamentally
different origin which stems from the self-generated active cell motion.
Moreover, it was noted that cell layers are governed by additional processes
such as cell division, apoptosis, and proliferation which are rigorously absent
in, e.g., colloidal glass-forming systems. 

In 2015, Garcia \textit{et al}.\ \cite{Garcia2015} studied the dynamics of a different
two-dimensional epithelial layer comprised of human bronchial epithelial cells
(HBEC). As in the study of Angelini and co-workers \cite{Angelini2011}, they found evidence for
dynamical heterogeneity and a significant dynamical slowdown with increasing cell
density. Interestingly, they also extracted a dynamical correlation length
associated with heterogeneous relaxation that changes non-monotonically with
the overall cell density. Furthermore, it was found that the cells exhibit
non-trivial correlations in their instantaneous velocities \cite{Garcia2015}. This finding is a
striking departure from conventional equilibrium physics: in non-active and non-driven
systems, the velocities of different particles are always uncorrelated \cite{Szamel2015a,Berthier2019}.
Curiously, the correlation length associated with the cells' velocities also
revealed a non-monotonic dependence on cell density, which in turn was
correlated with the length scale of dynamical heterogeneities \cite{Garcia2015}. Thus, the
underlying physics of the observed cellular glassy dynamics, in particular the
emergence of dynamical heterogeneity, may be fundamentally distinct from the
non-active equilibrium case. Garcia \textit{et al}.\ also concluded that the
cellular monolayer exhibits features of aging which are not merely due to a
gradual increase in cell density, but rather arise from the maturation of
cell-cell and cell-substrate contacts \cite{Garcia2015}. This result again constitutes a unique
aspect of living cells that has no counterpart in ordinary glassy materials. 

The relevance of collective cellular glassiness for pathological conditions
such as asthma was first realized in 2015 by Park \textit{et al}.\ \cite{Park2015}. By studying
confluent HBEC layers, they found that the cell dynamics of non-asthmatic
(i.e.\ healthy) donors underwent a continuous transition from a mobile,
fluid-like state to a quiescent, glass-like state as the cell layer matured
over time. Conversely, cells from asthmatic donors remained significantly more
mobile and exhibited a delayed glass-like transition. They also found that both
cell cultures were dynamically heterogeneous, with maximum correlated cluster
sizes on the order of 20 cells. Importantly, they identified a dimensionless
metric for the average cell shape--specifically, the average ratio $\bar{p}$
between the cell perimeter and the square root of the cell area--as a
\textit{structural} signature of glassiness. More explicitly, a $\bar{p}$ value
above 3.81 corresponds to fluid-like behavior, while $\bar{p} < 3.81$ indicates
rigidity of cell layer (also see Sec.\ \ref{sec:simcells}). Note that the cell-shape index $\bar{p}$ is inherently
a \textit{many-body} property; this must be contrasted with conventional
two-particle structural quantities of glass-forming liquids such as the radial
distribution function or static structure factor \cite{Hansen2013}. It remains to be established
whether two-body liquid-state-theory concepts of (active or
passive) particles will also be meaningful to describe the properties of
confluent and deformable living cells. One of the first steps in this direction was recently
taken by Giavazzi \textit{et al.}\ \cite{Giavazzi2017} to relate the static structure factor  
of cell nuclei to the flocking behavior of an epithelial monolayer. 

It was recently recognized that glassy dynamics is also manifested in
\textit{three-dimensional} cellular collectives, with relevance in both healthy
and diseased tissues. Indeed, processes such as wound healing, embryonic
development, and cancer are
all governed by transitions between migratory ('liquid-like') and stationary
('glass-like') cellular states \cite{Szabo2010,Kabla2012,Park2016,Hakim2017,Oswald2017,Malinverno2017,Palamidessi2018}. 
For example, in the context of embryonic
development, Sch\"{o}tz \textit{et al}.\ \cite{Schotz2013a} analyzed individual cell tracks of
three-dimensional tissue explants from zebrafish embryos. They found that the
cell dynamics inside the tissue exhibits subdiffusive and caging behavior
which, based on a minimal model, they attributed to enhanced cell-cell adhesion
and decreased active force generation. Another potentially important role of
glassy dynamics lies in the pathology of cancer: works by e.g.\ Friedl and
co-workers \cite{Haeger2014}, Park \textit{et al}.\ \cite{Park2016}, Oswald \textit{et al}.\ \cite{Oswald2017}, and
Palamidessi \textit{et al}.\ \cite{Palamidessi2018} hypothesize that
cancer metastasis is governed by cell 'unjamming' behavior that allows clusters
of cells to mobilize and ultimately escape the solid primary tumor. It is not implausible 
that glassy dynamical heterogeneities may play a role in this process \cite{Park2016,Giuliano2018}.
Recent studies by Malinverno \textit{et al.}\ \cite{Malinverno2017} and Palamidessi \textit{et al.}\ \cite{Palamidessi2018}
have further identified a molecular pathway toward fluidization of two- and three-dimensional 
kinetically arrested cellular collectives, revealing that overexpression of a protein called RAB5A
is sufficient to initiate collective cell motion. 
However, as in many other cell studies, biological factors such as varying
cell-cell adhesive contacts, genetic heterogeneities, the extracellular matrix
environment, and the cells' persistent active motion must also be considered to
ultimately understand the fate of a cellular collective; this is again a
fundamental difference with the description of 'simple' glassy liquids.

\subsubsection{Summarizing remarks}
In summary, there is now compelling evidence that living active systems exhibit
several hallmarks of glassy dynamics, both at the intra- and intercellular
level, both in two and three dimensions, and both in healthy and diseased tissue.
Importantly, this manifestation of glassiness may also carry a biological
function, e.g.\ by enabling individual cells to switch into a dormant glassy
state under unfavorable environmental conditions, and by allowing collectives
of cells to switch between quiescent and migratory behavior to facilitate
multicellular processes such as wound healing and tissue development.
Conversely, the apparent lack of cellular glassiness may underlie pathological
conditions such as asthma and cancer metastasis. Furthermore, due to the cells'
innate activity, and the fact that such activity can influence the glassy
dynamics, it is plausible that nature also employs activity as a means to
\textit{control} the cellular glassiness. For example, in collective cell
movement, it might be possible that the fragility of a cell sheet--i.e.\ the
abruptness with which a migrating cell layer comes to a halt--is tuned and
controlled by the cells' own self-propulsion forces. In the context of cancer,
the possible active inhibition of dynamically heterogeneous regions within the
primary tumor may potentially suppress the emergence of metastasizing cell
clusters. Future work will hopefully shed more light on these speculative but
highly interesting directions of research.

\section{Numerical simulation studies of active glassy matter}
\label{sec:sim}

In addition to experiments, computer simulations constitute a powerful
complementary approach to study complex physical phenomena such as glass
formation. These "numerical experiments" allow for controllable tuning of the
relevant parameter space whilst providing detailed particle-resolved
information on the emergent structural and dynamical properties. It is
therefore no surprise that simulations are now also widely employed to
elucidate the behavior of active glassy matter. In this section we review some
of the main simulation studies in this field, both for synthetic and living
model systems.

\subsection{Synthetic active glasses}
\label{sec:simsynth}

\begin{figure}
        \begin{center}
    \includegraphics[width=0.48\textwidth]{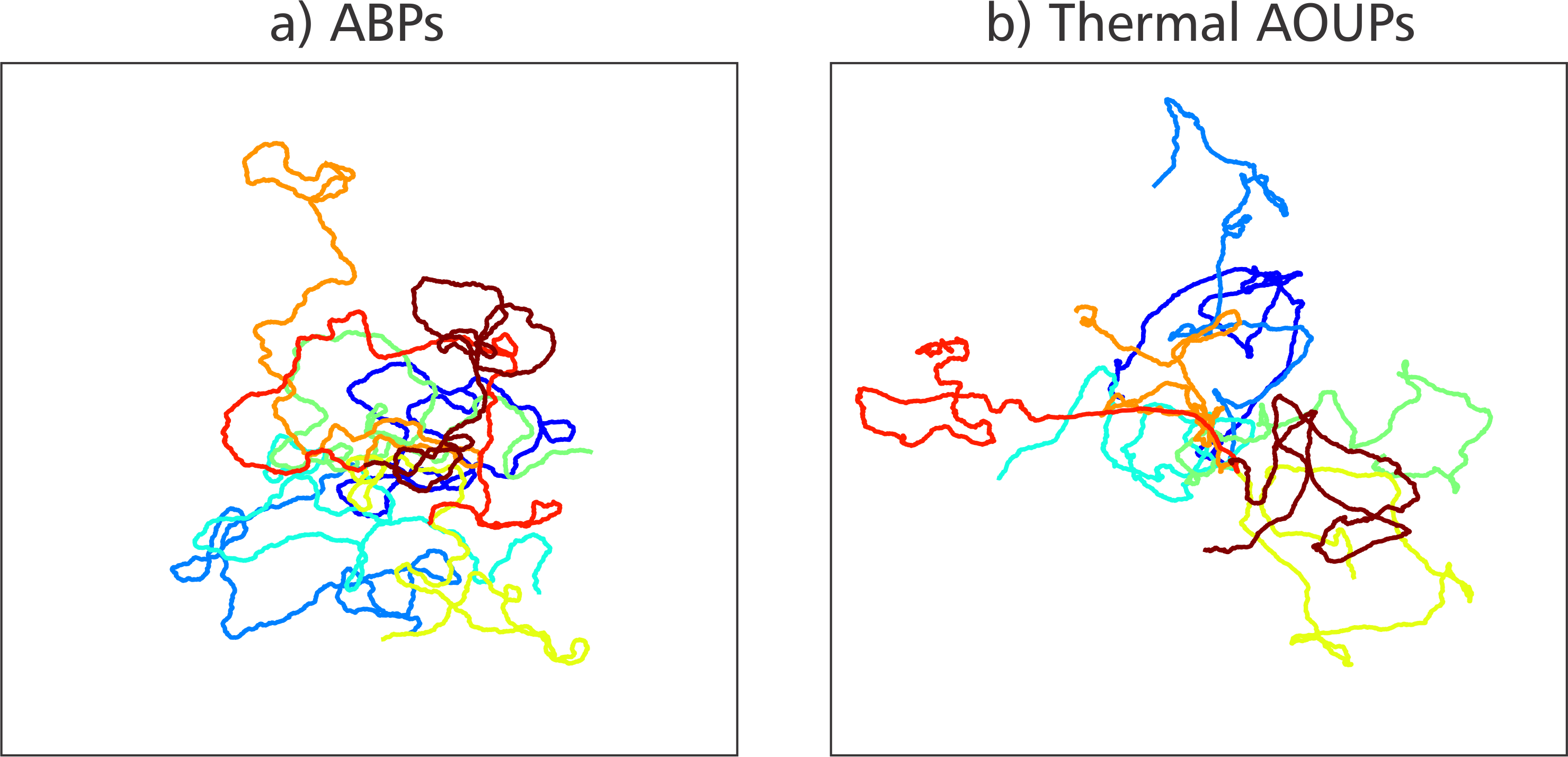}
  \end{center}
  \caption{
  \label{fig:ABPvsAOUP}
  Typical two-dimensional trajectories of (a) Active Brownian Particles and (b) thermal Active Ornstein-Uhlenbeck Particles.
  Each curve represents an independent trajectory for a non-interacting particle. 
  The model parameters were chosen such that the mean-squared displacements of both systems are identical.
  }
\end{figure}

For the study of synthetic active glassy materials, there now exist two popular
classes of simulation models: Active Brownian Particles (ABPs) \cite{Romanczuk2012} 
and Active
Ornstein-Uhlenbeck Particles (AOUPs) (see Fig.\ \ref{fig:ABPvsAOUP}) \cite{Szamel2014,Maggi2015}. 
Aside from particle interactions, the translational motion of
ABPs is governed by thermal Brownian motion and a constant self-propulsion
speed; the directionality of the self-propelled motion is controlled by
Brownian rotational diffusion. The active motion of AOUPs, on the other hand,
is governed by the Ornstein-Uhlenbeck process, which provides a stochastic
element to both the magnitude and direction of the self-propulsion forces. This
culminates into a persistent random walk that is characterized by a persistence
time--which describes the duration of persistent active motion--, and an
effective temperature which quantifies the strength of the active forces.  For
so-called athermal AOUPs, the effective temperature provides the only source of
motion (aside from particle-particle interaction forces) \cite{Szamel2015a}; for thermal AOUPs,
the translational motion is additionally governed by Brownian fluctuations \cite{Feng2017}.
Both ABP and AOUP systems are typically considered in the overdamped limit,
i.e.\ neglecting inertial effects, but underdamped dynamics may in principle
also be included (see, e.g., \cite{Cugliandolo2017}). In the limit of vanishing
self-propulsion speed or vanishing persistence time, respectively, both the ABP
and AOUP model reduce to a passive Brownian system.

\subsubsection{Dynamical slowdown, fragility, and dynamical heterogeneity}
\label{sec:simsABPOUP}

The 2013 study by Ni, Cohen Stuart, and Dijkstra \cite{Ni2013} constitutes the first
active-matter simulation of arguably the simplest structural glass former,
colloidal hard spheres.  They considered a dense system of ABPs in three
dimensions, and made several interesting observations regarding their glassy
dynamics. It was found that the structural relaxation time decreases
monotonically with increasing activity, implying that enhanced self-propulsion
leads to the effective breaking of cages and more liquid-like behavior. In fact,
they found that increased activity can ultimately push the glass transition
density toward random close packing, i.e.\ the highest possible density for
disordered spheres. A 2014 study by Wysocki \textit{et al}.\ \cite{Wysocki2014} on
three-dimensional soft active particles also reported liquid-like behavior
up to densities near random close packing. 
In terms of the fragility, Ni \textit{et al}.\ \cite{Ni2013} found that a higher
self-propulsion speed makes the system more fragile, corresponding to a more
abrupt vitrification process upon compression.  Features of dynamical
heterogeneity were also observed; at a fixed density, however, enhanced
activity was found to decrease the cooperative motion, concomitant to the
overall decrease in relaxation time. Finally, Ni and co-workers analyzed the
changes in microstructure at the level of the static structure factor. They
found that increasing the self-propulsion strength leads to a smaller average
nearest-neighbor distance but a broader distribution of interparticle
distances. They thus concluded that the structure of an equilibrium hard-sphere
glass is different from a non-equilibrium active hard-sphere system at the same
packing fraction. However, it was hypothesized that the microstructures should
become similar in the jamming limit at random close packing. A later study 
by De Macedo Biniossek \textit{et al.}\ \cite{DeMacedoBiniossek2018} on two-dimensional 
ABP hard disks confirmed that activity induces non-trivial microstructural changes,
which they attributed to a competition between activity-induced fluidization and 
enhanced structural order. 

Around the same time as the study by Ni \textit{et al.}, Berthier employed \cite{Berthier2014}
Monte Carlo simulations to explore the glassy dynamics of two-dimensional 
AOUP-like hard disks. While the governing equations for this
system are somewhat different from AOUPs, the particles also undergo a
persistent random walk in the dilute limit. Berthier found signatures of caging
and ultimate kinetic arrest as the particle density increases, as well as
dynamically heterogeneous correlated motion, reminiscent of conventional glass
phenomenology. However, the competition between caging behavior and active
persistent motion led to quantitative differences with respect to the
equilibrium case. Similar to the study of Ni \textit{et al}.\ \cite{Ni2013}, increased activity
was found to monotonically shift the glass transition density to a higher
value, implying that non-equilibrium self-propulsion indeed opens up new
relaxation pathways that are closed in equilibrium. It was suggested that this
might be due to an 'effective' attractive force that emerges as active particles
propel into each other, making the observed behavior of repulsive active
disks somewhat similar to the equilibrium behavior of adhesive hard spheres.
Indeed, it is well established that the addition of short-ranged attractions 
(through e.g.\ depletion interactions) can fluidize a passive hard-sphere glass \cite{Pham2002};
the work of Berthier \cite{Berthier2014} suggests that activity may possibly play a similar role as effective depletion. 

Later simulations by Szamel, Flenner, and Berthier \cite{Szamel2015a,Flenner2016} provided a more detailed
picture of the glassy properties of (athermal) AOUP systems. They focused on an
active version of the three-dimensional Kob-Andersen Lennard-Jones mixture--a
well-studied benchmark system for passive glass formation. It was found that,
contrary to the behavior of ABPs, active OUPs can become both more liquid-like
and more glass-like depending on the nature of active motion. Specifically, for
small persistence times at a fixed effective temperature, the active fluid
relaxes faster than a passive reference system, but for large persistence times
the active material relaxes more slowly. Interestingly, this non-monotonic
dependence of the relaxation time was accompanied by a \textit{monotonic}
growth of the main peak of the radial distribution function, implying that the
structure-dynamics link in active glassy OUPs is even more complex than in
equilibrium supercooled liquids. To account for the non-monotonic behavior,
they identified another important feature of the glassy dynamics which is
unique to non-equilibrium systems, namely non-zero correlations among
instantaneous particle velocities.  The existence of such non-trivial velocity
correlations  was already noted by Garcia \textit{et al.}\ \cite{Garcia2015} in an experimental
study on confluent cell layers, but Szamel, Flenner, and Berthier were the
first to recognize their importance in a combined numerical and theoretical
study of an active model glass former. More specifically, Szamel \textit{et
al.}\ \cite{Szamel2015a} argued that the non-monotonic change in the AOUP relaxation time with
increased activity is due to a competition between increasing velocity
correlations (which speed up the dynamics) and increasing structural
correlations (which slow down the dynamics). It was also found that the
fragility increases with increasing persistence time \cite{Flenner2016}, a result that is
consistent with the ABP results of Ni \textit{et al}.\ \cite{Ni2013}.  Finally, signatures of
dynamical heterogeneity and Stokes-Einstein violation were also observed.  After
rescaling all data with respect to the high-temperature behavior of the active
liquid, both dynamically heterogenous motion and Stokes-Einstein decoupling
were found to be rather similar to the equilibrium behavior in non-active
supercooled liquids \cite{Flenner2016}. 

A follow-up study by Berthier, Flenner, and Szamel on repulsive
(Weeks-Chandler-Andersen) AOUPs \cite{Berthier2017} suggested a more subtle
scenario of the interplay between glassy relaxation dynamics, microstructure,
and activity-induced velocity correlations. More specifically, they found that
the \textit{second} peak of the radial distribution function changes less
trivially than the first peak with increased activity, and it was argued that
these microstructural differences predominantly underlie the observed
non-monotonic changes in the relaxation time. Interestingly, a careful numerical
analysis showed that the velocity correlations grow only monotonically with the
persistence time. It was therefore concluded that, while emergent velocity
correlations \textit{accompany} the non-equilibrium glass transition, and
presumably affect the quantitative dynamics, they are not the main factor
responsible for the initial acceleration and subsequent slowdown of AOUP glassy
dynamics upon departing from equilibrium. It still remains to be established
whether this finding is universal or unique to the specific AOUP model, and whether
these insights may also be applicable to the glassy behavior of confluent active cells.  

A 2016 study by Mandal \textit{et al}.\ \cite{Mandal2016} focused on the glassy dynamics of
\textit{mixtures} of active and passive particles. They considered a
three-dimensional Kob-Andersen binary mixture in which only the minority (20\%)
species undergo active persistent motion; the direction of the active motion
was discretized in 8 directions. It was found that increasing the magnitude of
the self-propulsion force shifts the glass transition monotonically to a lower
temperature, akin to the monotonic increase of the glass transition density in
ABP hard spheres \cite{Ni2013}. Furthermore, they observed an activity-dependent
change in the fragility of the system, including a fragile-to-strong crossover
at sufficiently high self-propulsion strengths.  In contrast to the ABP results
of Ni \textit{et al}.\ and the AOUP simulations of Berthier \textit{et al.},
however, the binary material was found to become \textit{less} fragile with
increased activity. While the interaction potentials, material composition, and
active equations of motion differ among these studies, the qualitative
difference in fragility is striking. It must be noted, however, that Mandal \textit{et al}.\ \cite{Mandal2016}
considered the fragility as a function of the temperature $T$, whereas Berthier, Flenner, and Szamel \cite{Flenner2016,Berthier2017}
used the \textit{effective} temperature $T_{eff}$ as the control parameter. The latter accounts
for the \textit{total} amount of injected energy (i.e., energy of both thermal and active origin), and $T_{eff}$ may therefore
be a more suitable control parameter when the activity strength exceeds the typical thermal energy $k_BT$. Future work should clarify
whether the effect of non-equilibrium activity on the fragility can be captured in a single
unifying description; thus far, however, it appears that the microscopic details of the
active material also play a non-trivial role. 

\subsubsection{Aging dynamics}
\label{sec:aging}
Finally, let us discuss the fifth important aspect of glassy dynamics, namely
non-equilibrium aging. As noted in Sec.\ \ref{sec:aging}, aging in conventional glassy
systems is understood as a gradual approach toward a deeper energy minimum in a
high-dimensional energy landscape.  Active materials, however, are governed by
a constant injection and dissipation of energy at the single-particle level,
rendering them non-Hamiltonian systems.  Hence, the potential (or free) energy
is generally not a useful metric to describe the behavior of active matter, and
the manifestation of aging in an active glass is \textit{a priori} far from
trivial.  

The first numerical evidence that waiting-time-dependent aging dynamics can,
indeed, also take place in active glassy systems was presented in 2017 \cite{Janssen2017}. This
work focused on athermal active repulsive rods with a constant self-propulsion
speed; unlike ABP systems, however, the reorientation dynamics in the system
arises from collision-induced torques rather than rotational Brownian motion.
For sufficiently high densities and sufficiently short rods, it was found that
the system freezes into a disordered and kinetically arrested state--i.e., an
active glass. By subsequently decreasing and increasing the density in a
periodic manner, the athermal active glass could be periodically melted and
revitrified, somewhat similar to applying a temperature-cycling protocol for
thermal glass formers \cite{Zhao2013,Ketov2015}. Interestingly, upon repeated application of such
melting-revitrification cycles, the average number of rearranging particles would decrease progressively, 
ultimately yielding a configuration in which
all relative particle motion has ceased. That is, the system has manifestly aged
toward a more stable state that remains solid even at lower densities. It is
important to note that the particles' innate activity is a crucial
ingredient for this aging behavior; since the model lacks any thermal noise by
construction, the only non-trivial source of motion stems from the activity.
Indeed, in the passive reference system, the only form of time-dependent motion
would be an instantaneous quench toward the nearest energy minimum. 

\begin{figure}
        \begin{center}
    \includegraphics[width=0.48\textwidth]{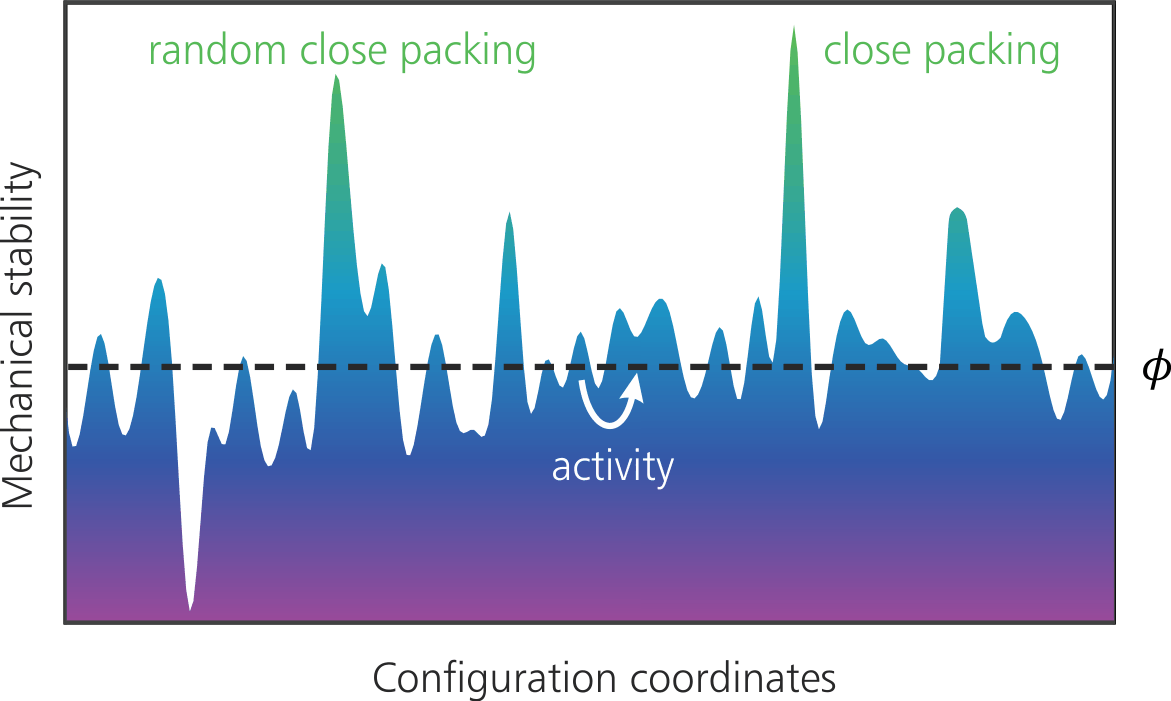}
  \end{center}
  \caption{
  \label{fig:stability}
  Illustration of the stability landscape picture. The $x$-axis
  represents all of coordinate space, and the $y$-axis represents the maximum density or packing fraction at
  which a configuration is mechanically stable, i.e., solid-like. The most dense configuration is the close-packed
  crystalline state, while the densest glass is at random close packing. Note that this example assumes that only a single crystalline state exists. 
The dashed line indicates a typical packing
  fraction $\phi$; in the absence of thermal fluctuations, non-equilibrium self-propulsion forces may allow an active system to surmount
  local barriers.
  The figure is adapted from Ref.\ \cite{Janssen2017}.  
  }
\end{figure}

To rationalize the observed aging behavior in active glasses, a new 'landscape'
paradigm was introduced which focuses not on the total (free) energy, but
rather on the \textit{mechanical stability} of the system \cite{Janssen2017}. Let us first recall
that aging in passive glass formers is governed by thermal fluctuations which
allow the system to surmount local energy barriers and ultimately reach a deep
energy minimum (Fig.\ \ref{fig:energy}).  For non-Hamiltonian active glasses, however,
Janssen, Kaiser, and L\"{o}wen \cite{Janssen2017} proposed that it is not the
energy that is optimized but rather the mechanical stability: as soon as the
active athermal material has reached a configuration that is sufficiently
stable to survive at the lowest applied densities, it will remain in that
configuration indefinitely. Such a state is characterized by a local force
balance on all particles; i.e., the intrinsic self-propulsion forces are
cancelled against the local particle-particle interactions to effectively stop
all relative particle motion. By subsequently plotting all possible global
stabilities of the system as a function of configuration space, a rugged
high-dimensional 'stability landscape' emerges (Fig.\ \ref{fig:stability}) that is reminiscent
of the conventional energy landscape picture of Fig.\ \ref{fig:energy}.  For the active
glassy system studied in Ref.\ \cite{Janssen2017},  the exploration of this
stability landscape is not facilitated by the presence of thermal fluctuations
but rather by active particle forces, and the observed aging behavior is
therefore a purely activity-induced phenomenon.  In analogy to work on
periodically driven systems \cite{Corte2008}, Janssen, Kaiser, and L\"{o}wen
\cite{Janssen2017} interpreted this novel form of active aging as an
irreversible, random self-organization process toward an 'absorbing state'.
That is, the active system continues to explore many different configurations
until it spontaneously reaches a stable state from which it can no longer
escape. We note that in principle there exist many such absorbing states for a
given range of densities; at the highest density, all possible absorbing states
should correspond to all (active) random close packing configurations.

The role of thermal noise in active glassy aging was also discussed in Ref.\
\cite{Janssen2017}. It was hypothesized that for thermal active glasses every absorbing state
on the stability landscape is effectively replaced by a \textit{basin} of
absorbing states, i.e.\ a set of similar configurations that are separated only
by relatively small barriers.  However, explicit aging simulations for such
thermal active systems have not yet been performed. In future studies, it will
be interesting to explore how active physical aging is affected by thermal
fluctuations, whether such fluctuations play the same effective role as innate
self-propulsion forces in the aging phenomenology of active glasses, and
whether e.g.\ temperature cycling and density cycling will amount to similar
aging behavior.  Finally, in the context of living systems, a phenomenological
link between physical aging of glasses and biological aging of cells was
recently established \cite{Lou2017}. This raises a highly intriguing question: can our
understanding of active glassy physics ultimately also reveal more insight into
the long-term aging behavior of living tissues?

\subsection{Living active glasses}

To simulate the dynamics of living cells, a multitude of coarse-grained
computer models have been developed, including simple isotropic and deformable
particle models, vertex and Voronoi models, lattice-based cellular Potts
models, and models with explicit subcellular elements \cite{Camley2017}.  Broadly speaking, these
respective approaches are characterized by an increasing level of complexity
but also by a growing computation cost.  Hence, there is \textit{a priori} not
a unique model that is ideally suited to describe the relevant physics of
living cell systems.  Here we focus only on simulation studies that
specifically address the glassy dynamics of confluent cell layers; owing to the
fact that such simulations require the explicit treatment of a large number of
individual cells, these studies have thus far been limited to relatively simple
descriptions at the single-cell level. For computational modeling efforts to
describe other biological phenomena such as (non-glassy) epithelial
morphogenesis,  we refer the reader to e.g.\ Ref.\ \cite{Fletcher2014}
and especially Ref.\ \cite{Camley2017}.

\subsubsection{Soft-particle models}
Arguably the simplest description of a glassy layer of active cells is a
soft-particle-based model, in which every cell is represented as a
self-propelled soft disk. This approach was first invoked by Henkes, Fily, and
Marchetti in a seminal 2011 study on dense active matter \cite{Henkes2011}. Their work in fact also
precedes any of the synthetic active-glass simulations mentioned in Sec.\ \ref{sec:simsynth},
and it can therefore be argued that this study constitutes the first numerical
realization of active glassy physics.  Briefly, Henkes \textit{et al.}\ modeled
a two-dimensional system of soft, isotropically-interacting repulsive disks
that are each equipped with a constant self-propulsion speed.  Inspired by
recent confluent-cell experiments \cite{Angelini2011, Angelini2010, Trepat2009, Poujade2007,Petitjean2010}, they included
an internal alignment mechanism such that the self-propulsion direction tends
to align with the particle's instantaneous velocity; in a biological context,
this mimics the effect of aligning the cell polarity with its motility.  By
calculating the state diagram for a circularly confined system, it was found
that the particles become kinetically arrested at sufficiently large densities and low
self-propulsion speeds, with a weak reentrant behavior at high degrees of
activity. We note that, while the authors referred to the arrested dynamics as 'active jamming', we here
follow the convention that this is an active glassy state, owing to the presence of active motion.  
 At the largest self-propulsion speeds studied, all trajectories
yielded fluid-like behavior.  Within the densely arrested phase, they also
observed regular oscillations of the particle displacements around their mean
caging positions, which--in analogy to athermal dense packings--they interpreted as low-frequency modes.

We note that the synthetic active-particle simulations discussed in Sec.\ \ref{sec:simsynth}
may, in principle, also be regarded as a minimal model for living cell tissues,
although these studies did not explicitly take into account any cell-specific
properties. In particular, the assumption of ABP- or AOUP-like dynamics may
conform more accurately to the behavior of artificial active colloids than to
densely packed living cells. Conversely, while the study by Henkes \textit{et
al}.\ assumed an equation of motion that was inspired by biological cell
polarization, their work may, of course, also shed light on the behavior of
synthetic materials with a similar active alignment mechanism.

\subsubsection{Vertex and Voronoi models}
\label{sec:simcells}

\begin{figure}
        \begin{center}
    \includegraphics[width=0.48\textwidth]{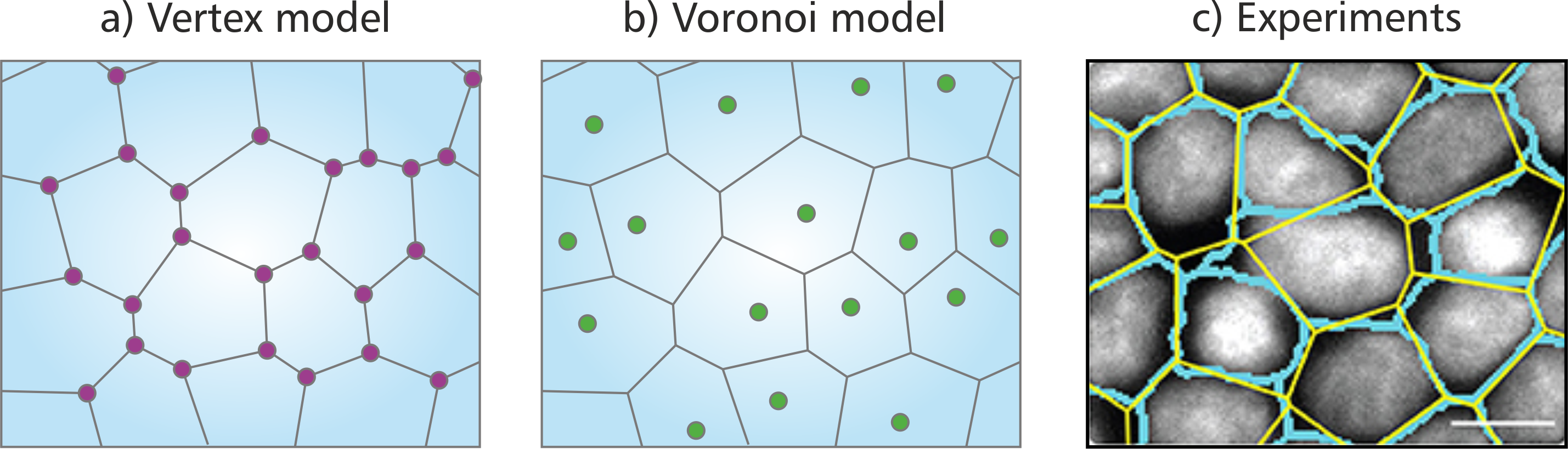}
  \end{center}
  \caption{
  \label{fig:vertexvoronoi}
  Illustration of (a) the vertex model and (b) the Voronoi model for confluent
cells. The purple and green circles represent the respective degrees of freedom
in these models.  Panel (c) shows the differences between the cell boundaries
of MDCK cells obtained via imaging methods (blue) and a Voronoi tesselation of
the cell nuclei (yellow). 
Figure (c) is reproduced from Ref.\ \cite{Kaliman2016}.
  }
\end{figure}

A more refined description of glassy cell sheets should also account for the
inherent confluence of the layer and the anisotropic shape deformability of the
cells.  These aspects are naturally incorporated in so-called vertex and
Voronoi models, which both represent cells as polygons that collectively
tesselate the entire space; as such, they are both confluent by construction.
The main difference between these models lies in the choice of the relevant
degrees of freedom: in vertex models, the dynamical equations of motion are
solved for the polygon vertices, while in Voronoi models the equations of
motion apply to the Voronoi centers of each cell (see Fig.\ \ref{fig:vertexvoronoi}). An advantage of the vertex
model is that it can describe both concave and convex cell shapes, while the
Voronoi model is inherently constrained to convex polygons. Nonetheless, it
must be noted that a Voronoi tesselation using the cell nuclei as Voronoi
centers can already provide a fairly accurate description of the cell shapes
observed in experiment \cite{Kaliman2016}.  For the incorporation of
active cell motility, the Voronoi model is generally preferred over a
vertex-model simulation, as the Voronoi centers provide a natural means to
assign a single active force to every individual cell.  Although the idea to
represent an epithelial cell layer by polygons was already proposed by Honda in
1978 \cite{Honda1978, Nagai2001}, it is largely due to the recent pioneering
work of Bi, Manning, and co-workers that such modeling approaches are now
recognized as a successful tool to describe and understand the glassy dynamics
of living cells.  We discuss some of these main simulation studies below.

In 2014 and 2015, Bi \textit{et al}.\ \cite{Bi2014, Bi2015} studied the glassy
dynamics of a confluent monolayer using a two-dimensional vertex model. Their
model assumed that the mechanical energy of each cell is governed by its
perimeter $P$ and surface area $A$, effectively accounting for the bulk
elasticity of the cell, the active contractility, and a net line tension. The
latter arises from a competition between the cortical tension of the active
actomyosin layer near the cortex--which tends to minimize the area of cell-cell
contact--, and intercellular adhesion forces which maximize the area of
cell-cell contact.  The cell shape can be non-dimensionalized in terms of the
so-called shape index $p=P/\sqrt{A}$; based on their simulations, it was
predicted that the confluent cell layer undergoes a fluid-to-solid-like
transition at an average value of $\bar{p}= 3.81$. This value could also be
associated with a change in the energy barrier heights of so-called T-1
transitions-- topological transitions which govern cell rearrangements in
confluent layers. 
 
The vertex-model-based prediction of a rigidity transition at $\bar{p}=
3.81$ was soon found to be in remarkable agreement with experiments of
confluent HBEC monolayers, as detailed in the 2015 study by Park \textit{et
al}.\ \cite{Park2015}.  In this joint experimental-simulation work, the effect
of intrinsic self-propelled cell motion on the transition was also studied. To
this end, a small active term was added to the model that, similar to the study
by Henkes \textit{et al}., was set to a constant magnitude and in a direction
that tends to align with the cell's velocity.  It was found that increased
active forces operate in concert with increased adhesion to fluidize the
cellular collective; however, the critical shape-index value at which the
transition takes place remains at $\bar{p}=3.81$, regardless of the magnitude
of the active fluctuations. It was also noted that increased cell-cell adhesion
leads to larger cell perimeters and more liquid-like behavior; paradoxically,
higher adhesion in particulate matter generally leads to enhanced gelation and
solidifaction, highlighting the fact that adhesive forces may play a
fundamentally different role in living and non-living glassy matter. We note, however, 
that short-ranged attractions in hard spheres can also fluidize a glassy state 
(and even give rise to reentrant behavior) \cite{Pham2002}; this may possibly point toward  
a non-trivial analogy between passive glass formers and active cellular systems.

\begin{figure}
        \begin{center}
    \includegraphics[width=0.48\textwidth]{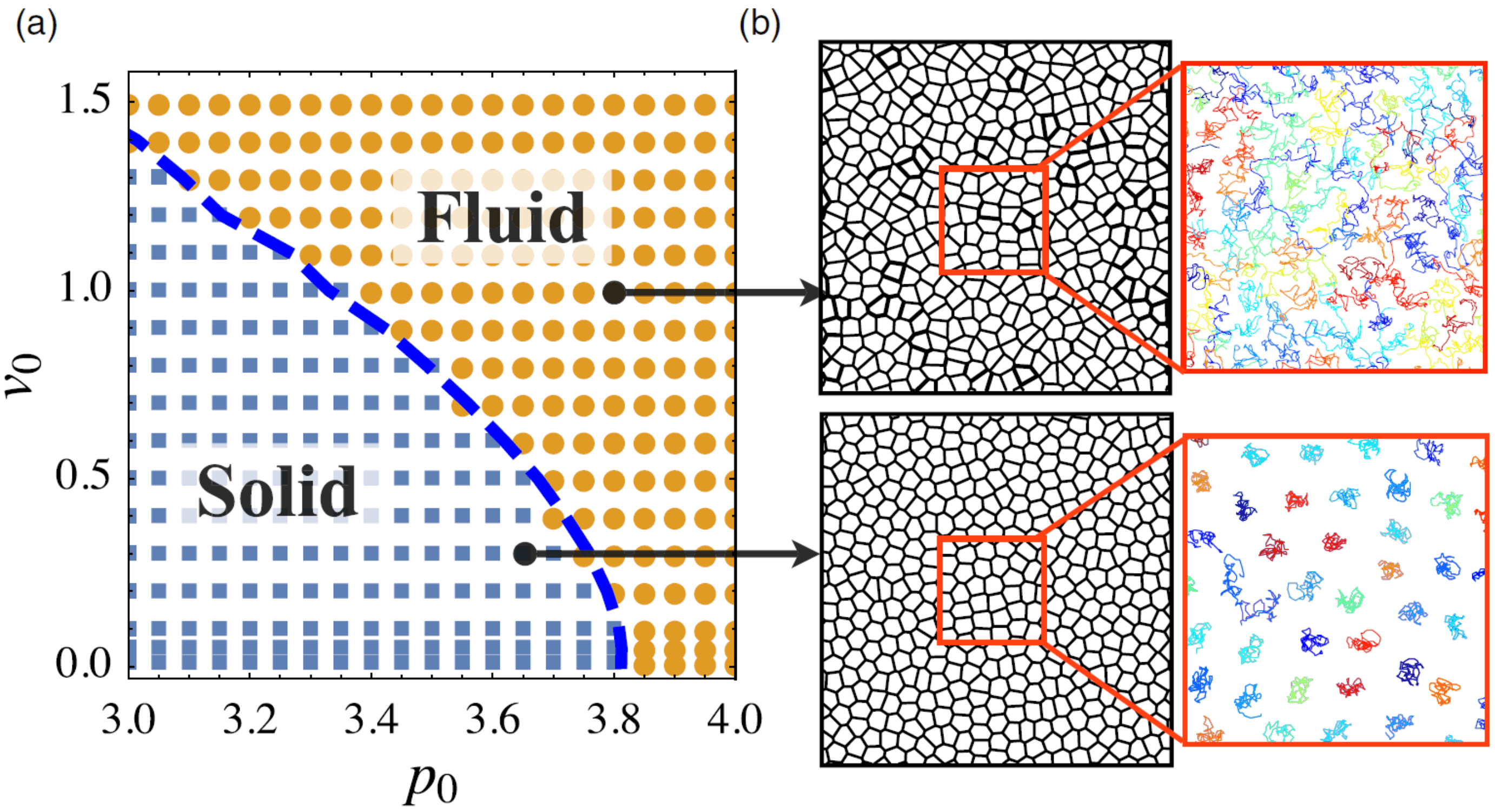}
  \end{center}
  \caption{
  \label{fig:SPV}
  (a) State diagram for the self-propelled Voronoi model as a function of the cells' self-propulsion speed $v_0$ and preferred cell perimeter $p_0$. 
  Blue data points indicate solid-like behavior (vanishing diffusion constant) and orange points indicate liquid-like behavior (finite
 diffusion constant). The blue dashed line corresponds to an average shape index of $\bar{p}=3.81$. 
  (b) Instantaneous snapshots show the difference in cell
shape across the transition. Cell tracks also show dynamical arrest due to caging in the solid phase and diffusion in the fluid phase.
The figure is reproduced from Ref.\ \cite{Bi2016}.
  }
\end{figure}

A 2016 simulation study by Bi \textit{et al.}\ \cite{Bi2016} focused on the
glassy dynamics of confluent cell layers using the so-called self-propelled
Voronoi model.  The mechanical energy of the system was assumed to be the same
as in the vertex model, i.e.\ controlled by the cells' perimeters and areas.
For the self-propulsion forces, an ABP-like equation of motion was invoked in
which the activity vector has a constant magnitude $v_0$ and a directionality
that is governed purely by Brownian rotation. The main result of this study is
the state diagram shown in Fig.\ \ref{fig:SPV}, which indicates regions of parameter space
in which the dynamics is glass-like and fluid-like, respectively. Notably,
while the degree of glassiness is explicitly and monotonically dependent on both
the magnitude of the self-propulsion forces and the preferred perimeter of the
cells, the point at which the glass transition takes place always corresponds
to an average shape-index value of $\bar{p}= 3.81$. The role of the rotational
diffusion constant was also explored, revealing a monotonic shift in the
glass-transition line.  More specifically, a high rotational diffusion constant
was found to give rise to largely uncorrelated motion and more solid-like
behavior, while a low rotational diffusion constant induces more collective
motion and flow within the tissue. In the limit of vanishing self-propulsion
speed, all glass-transition lines converge to the point at which the preferred
cell perimeter $p_0$ equals $\bar{p}=3.81$. A later study by Merkel and Manning
\cite{Merkel2018} revealed that the rigidity transition of the Voronoi model in
three dimensions is similarly controlled by the surface-to-volume ratio of the
cells.  In view of these findings, and their remarkable agreement with earlier
vertex-model and experimental studies of glassy cell layers, it appears evident
that there is a clear and quantitative correlation between the microstructural
cell shapes  and the glassy dynamics of a cellular collective.  As already
noted in Sec.\ \ref{sec:exptlife}, however, the structural role of the shape index may be
unique to confluent deformable cells and need not necessarily apply to
isotropic particles undergoing a glass transition, thus possibly constituting a
fundamental difference between the glassy physics of living and non-living
matter.

A more recent and larger-scale simulation study by Sussman and Merkel \cite{Sussman2018a} on the
self-propelled Voronoi model highlights the difference between glassy and
jamming physics in confluent models. In particular, they found that the
zero-motility limit ($v_0=0$) of the two-dimensional self-propelled Voronoi
model is always marginally stable, i.e.\ the number of particle constraints and
the number of degrees of freedom are exactly balanced, regardless of the value
of the shape perimeter $p_0$. Consequently, although the non-motile Voronoi
system undergoes an apparent fluid-to-solid glass transition at $p_0 = 3.81$,
the system strictly lacks an unjamming transition.  The two-dimensional vertex
model \cite{Bi2015} and three-dimensional Voronoi model \cite{Merkel2018}, on
the other hand, are always under-constrained and are rigidified by the
collective onset of residual stresses. Overall, the work of Sussman and Merkel
suggests a scenario in which the dynamical glass transition in the
two-dimensional $v_0=0$ Voronoi model is governed by the time scale of
rotational diffusion rather than by the shape of the cells, and points toward a
non-trivial decoupling of glassy and jamming phenomena.

The first simulation study on dynamical heterogeneity and fragility in confluent
glassy models was presented in 2018 by Sussman \textit{et al}.\
\cite{Sussman2018}. They studied the glassy dynamics of a two-dimensional
\textit{inactive} variant of the Voronoi model, in which the translational motion of a
cell is governed solely by cell-shape-based interactions and thermal Brownian
noise.  They found that the dynamics slows down significantly as the
temperature is reduced; however, it was noted that the plateauing region in the dynamical
scattering function, which is typically associated with caging, is much less
pronounced than in a standard glass former. At the corresponding time scales of
relaxation, a growing peak of the dynamical susceptibility was observed,
indicating marked dynamical heterogeneity and increased cooperative particle
motion. Interestingly, in contrast to the self-propelled Voronoi model, they
found that the location of the glass transition does not coincide with an
average cell shape index of $\bar{p}=3.81$. Thus, for this particular inactive
system, the dynamics and microstructural properties are manifestly decoupled.
Furthermore, they found an anomalous pattern in the fragility: as the
temperature decreases toward the glass transition temperature, the relaxation
time grows in a \textit{sub-Arrhenius} fashion. In terms of the conventional
energy-landscape picture, such 'superstrong' behavior would correspond to a
progressive decrease in local energy barriers as the system is cooled.
Similar qualitative results were found for the inactive vertex model. It
remains to be established whether this unusual fragility is unique to the
specific choice of the model, and whether the inclusion of active forces would
yield a different phenomenology. It must also be noted that experiments by
Angelini \textit{et al}.\ on epithelial MDCK layers showed a moderately fragile
behavior upon increased cell density \cite{Angelini2011}; it is not yet clear how the here reported
simulation results can be applied to those experiments, and whether superstrong
fragilities indeed may be observed in real biological tissues. 

We conclude this section by mentioning several other, recently developed
variants of vertex and Voronoi-based simulation models of confluent cells.
Giavazzi \textit{et al}.\ \cite{Giavazzi2018} studied a self-propelled Voronoi model in which the
self-propulsion direction is not governed by Brownian motion, but is rather
enslaved to the cell's own velocity. It was found that, within this model, such
polarization alignment yields global migration and can promote both dynamical
heterogeneity and solidication. L{\aa}ng and co-workers \cite{Lang2018} recently developed a
new model which combines the active Voronoi model with the Vicsek model, such
that every cell's self-propulsion direction tends to align with the
instantaneous velocity of its neighbors.  This model predicts large-scale
collective motion as the Vicsek-alignment radius reaches a certain threshold,
in good agreement with experiments on confluent human keratinocyte cells. On a
phenomenological level, the growing Vicsek radius captures the effect of
up-regulating cell-cell adhesion by experimentally increasing the calcium
concentration; the activity could be controlled experimentally by adding blood
serum to the cell layer \cite{Lang2018}.  Barton \textit{et al}.\ \cite{Barton2017} studied an active vertex
model which--in addition to cell-specific mechanical properties and cell
alignment--also accounts for cell growth, division, and apoptosis. It was found
that fluidization in this model can be induced either by increasing the
preferred cell perimeter $p_0$ or the magnitude of the active driving. Very
recently, Czajkowski \textit{et al}.\ \cite{Czajkowski2019} also simulated an
active vertex model that includes cell motility, cell division, and cell death.
They found that a glass-like regime with caging behavior and subdiffusive
dynamics can be achieved if the rate of cell cycling is sufficiently low.
Finally, a recent series of papers by Ruscher and co-workers \cite{Ruscher2015,Ruscher2017,Ruscher2018} studied the glassy
physics of the so-called Voronoi liquid, which is defined such that each
Voronoi position tends to move towards the centroid of the Voronoi cell to
which it belongs. Although this model was not motivated by biological
experiments, it revealed a multitude of non-trivial glassy features; it will be
interesting to explore if and how the Voronoi liquid can be linked to other
Voronoi-based models of confluent living cells.

\subsubsection{Cellular Potts models}
Let us finally mention one other model used for cell-resolved simulations, namely the
cellular Potts model. Briefly, this coarse-grained model represents each cell as a domain of connected spins
on a lattice; the dynamics of the cell is governed by an effective energy functional which captures
all relevant cellular interactions. In 2016, Chiang and Marenduzzo \cite{Chiang2016} developed an active variant
of the confluent cellular Potts model that is reminiscent of the vertex model of Bi and co-workers \cite{Bi2015,Park2015}.
They found that, depending on the cell motility and the interfacial tension between adjacent cells, the model can exhibit 
several features of glassy dynamics, including very slow relaxation, subdiffusive behavior, and aging.
Interestingly, it was found that the fluid-to-solid transition in the cellular Potts model also coincides rather accurately with a specific
average shape index $\bar{p}$. This suggests that predictions of the vertex and Voronoi model, as well as observations in experiment,
may also be captured within a lattice-based cellular Potts model.  

\subsubsection{Summarizing remarks}
In summary, there now exist numerous simulation models that can provide new
insight into the glassy dynamics of densely packed cell layers, in particular
on the emergence of a dynamical slowdown and dynamically heterogeneous
behavior, as well as on the relevant underlying microstructural changes of the
tissue. Thus far, based on several coarse-grained model studies, it
appears that a suitable means to capture the structural signature of glassy
behavior is encoded in the geometric shape of the cells, although this result
need not apply generally to all simulation models of cellular collectives.
Furthermore, it is not yet clear whether alternative structural metrics, such
as the radial distribution function used in liquid-state theory, may be
successfully applied to biological tissues.  Aside from the question to what
extent the above mentioned simplified cell models can capture all the relevant
physics of a realistic biological system, there are also still major unresolved
issues relating to the main hallmarks of glassy dynamics in living systems.
Notably, the degree of fragility in a confluent cell layer, the manifestation  
of Stokes-Einstein violation, the physical origins of aging behavior, as well as the possible role
of intrinsic cell motility on all these phenomena, still remain to be explored.

\section{Theoretical studies of active glassy matter}
\label{sec:theory}

Theoretical physics offers a third important methodology to study the glassy
phenomenology of active matter. During the last decades, a wide variety of
theories has been developed for 'passive' glass formation, including Adam-Gibbs
theory \cite{Adam1965}, spin-glass theory \cite{Castellani2005}, Mode-Coupling Theory (MCT) \cite{Gotze2008}, Random First Order
Transition Theory (RFOT) \cite{Lubchenko2007}, geometric frustration \cite{Tarjus2005}, dynamical facilitation \cite{Chandler2010}, and
kinetically constrained models \cite{Cancrini2009}. The differences between these theories
essentially stem from their physical point of view, which may be purely
thermodynamic, purely dynamic, purely phenomenological, or a combination
thereof.  For a detailed overview of existing theories of the glass transition,
see e.g.\ Refs.\ \cite{Binder2011, Berthier2011a,Langer2014, Royall2015}. In this section, we review several recent studies which
have sought to extend the theoretical description of glass formation to
non-equilibrium active systems. In particular, we focus on studies of spin
glasses (i.e., disordered spins on a lattice) and active variants of both MCT
and RFOT.  These approaches have been developed for model glass-forming systems
that are governed by a certain effective Hamiltonian (in the case of active
spin glasses), or by ABP- and AOUP-like dynamics (in the case of active MCT and
RFOT). As such, they do not distinguish \textit{a priori} between synthetic and
living active materials, but rather give general predictions for systems that
conform to certain equations of motion. 

\subsection{Spin-glass theories}
The first theoretical description of active glassy physics is the seminal 2013
study by Berthier and Kurchan \cite{Berthier2013}. They studied the mean-field behavior of a
non-equilibrium variant of the so-called spherical $p$-spin model; in the
equilibrium case, this spin-glass model can be solved exactly and conforms to
the RFOT scenario \cite{Kirkpatrick1987}.  To make the model active, they
included a colored-noise driving term and a dissipative memory term in the
effective
Hamiltonian.  It was found that both colored noise and energy dissipation give
rise to a dynamical slowdown that is strongly reminiscent of equilibrium glass
formation. The relationship between the magnitude of the driving force and the
decrease in the glass transition temperature was found to be roughly linear.
Overall they concluded that, although the specific features of the transition
may change quantitatively with the nature of the non-thermal driving force, the
phenomenology of the slowdown is rather similar to the behavior of thermal
glass-forming systems. This suggests that non-equilibrium active glassy systems
may, at least in approximate manner, be mapped onto an effective equilibrium
material. 

In 2014, Pilkiewicz and Eaves \cite{Pilkiewicz2014} analyzed the glassy properties of an active
generalization of the Sherrington-Kirkpatrick model--another well-known
spin-glass model.  Specifically, they considered a mixture of passive and
active spins, in which the passive spins possess quenched (fixed) disorder
while the active spins can undergo annealing (i.e., relaxation of the spin
orientation) due to an external driving force.  Upon increasing the fraction of
active spins, the glass transition temperature was found to shift to lower
values in a monotonic manner. However, when a bias was introduced to promote
ferromagnetic order, the phase diagram became markedly more complex: for a
certain range of the bias parameter, a reentrant behavior was predicted in
which the system goes respectively from a paramagnetic ('liquid-like') phase at
high temperatures, to a spin glass at lower temperatures, to a ferromagnetic
('crystal'), and finally to a mixed spin-glass phase with partial order at the
lowest possible temperatures.  Although the inherent lack of microstructure in
spin-glass models renders them, in principle, unsuitable to describe realistic
structural glass-formers, it was suggested that the bias parameter may possibly
be mapped onto the bulk density of an experimental mixture of active and
passive colloids.  Thus far, the glassy dynamics of such active-passive
mixtures still remains to be explored in experiment.  

\subsection{Mode-coupling theories}
We now turn to another important class of theories, namely Mode-Coupling
Theory \cite{Gotze2008,Reichman2005,Janssen2018}. MCT is essentially the only theory of the glass transition that is
founded entirely on first principles, aiming to predict the full microscopic
glassy dynamics of a material based solely on knowledge of the material's
microstructure.  Briefly, the theory starts from the \textit{exact} equations
of motion for a correlated (glassy) liquid, and subsequently applies a series
of approximations to obtain a closed equation for the dynamics; this equation
can be solved once the static structure factor is provided as structural input.
Although MCT in its standard form cannot adequately account for all hallmarks
of glassy dynamics, the theory has nonetheless provided a remarkable series of
non-trivial predictions for the qualitative and quantitative dynamical slowdown
of many glass-forming materials, including the existence of the cage effect,
non-trivial scaling laws, non-exponential relaxation, and complex reentrant
behavior. Furthermore, proposed modifications of standard MCT may offer more
accurate first-principles predictions of, e.g., fragility \cite{Janssen2014,Janssen2015a},
dynamical heterogeneity \cite{Biroli2006}, Stokes-Einstein violation
\cite{Charbonneau2018}, and aging \cite{Latz2000,Ramirez2010a}.  We refer the reader to
Refs.\ \cite{Gotze2008,Reichman2005,Janssen2018} for a general overview of the
technical aspects, successes, failures, and proposed extensions of standard
MCT. Here we focus only on new variants of MCT developed specifically for
active matter; a review of this emerging field was also recently presented in
Ref.\ \cite{Janssen2018}, and we summarize only the key findings below. 

Farage and Brader were the first to extend the standard MCT equations to an
active-matter system \cite{Farage2014}. In their 2014 paper, they considered the glassy dynamics
of hard-sphere ABPs by treating every active particle as an effectively
\textit{passive} colloid but with a higher, activity-controlled diffusion
constant. This approach essentially amounts to the explicit removal of all
rotational degrees of freedom.  Within this approximation, it was found that
the increase of particle activity can enhance the breaking of local cages,
resulting in a softening and ultimate melting of a passive glass. The location
of the glass transition was found to shift monotonically toward higher
densities, consistent with the ABP computer simulations of, e.g., Ni, Cohen
Stuart, and Dijkstra \cite{Ni2013}.  A similar MCT approach was later used by
Ding \textit{et al}.\ \cite{Ding2015} to study the dynamics of mixtures of hard-sphere active
and passive particles.  They predicted a reentrant behavior in the glassy
dynamics: upon increasing the self-propulsion strength of the active fraction
at constant density, the system changes first from a glass to a liquid and then
to a glass again. It was argued that this reentrance is unique to
active-passive mixtures and would disappear in a purely active system. Note,
however, that the here reported reentrance is fundamentally distinct from the
spin-glass results of Pilkiewicz and Eaves \cite{Pilkiewicz2014}.

In 2015, Szamel and co-workers presented the first microscopic MCT framework
for athermal AOUP systems \cite{Szamel2015a, Szamel2016}.  By assuming that the
particle positions evolve on a time scale much larger than the time scale
needed for reorientation of the activity direction, the self-propulsion could
be integrated out in an approximation akin to the one invoked by Farage and
Brader.  An important difference with earlier MCT work, however, is that
Szamel's theory also requires static correlations between particle velocities
as input. Contrary to the behavior of ABPs, and in remarkable agreement with
AOUP computer simulations, it was predicted that the incorporation of activity
can both enhance and suppress glass formation, depending on the persistence
time of the active particles. As already noted in Sec.\ \ref{sec:simsABPOUP}, they attributed
this non-monotonic behavior to a competition between increasing velocity
correlations and increasing structural correlations. For sufficiently large
persistence times, it was found that the fitted MCT glass transition
temperature increases monotonically with increasing persistence time,
suggesting that vitrification occurs more easily as the material becomes more
active.  An MCT-based scaling analysis for this type of active-matter system
was later performed by Nandi and Gov \cite{Nandi2017a}.

Feng and Hou subsequently studied a \textit{thermal} version of the active
Ornstein-Uhlenbeck model that also includes thermal translational noise \cite{Feng2017}. Their
MCT derivation differs fundamentally from the approach taken by Szamel \cite{Szamel2016},
however: it is valid only for sufficiently small persistence times and does not
require explicit velocity correlations as input.  Instead, their active-MCT
dynamics is governed by an averaged diffusion constant and a steady-state
structure factor which both depend on the effective temperature, density, and
the persistence time of the active particles.  They found that the critical
glass-transition density shifts monotonically to larger values with increasing
magnitude of the self-propulsion force, and that the critical glass-transition
temperature increases with the persistence time. These results highlight the
fact that different approximations within the theory can lead to fundamentally
different results, and that thermal fluctuations may play a non-trivial role in
the glassy dynamics of active Ornstein-Uhlenbeck particles. In a more recent study, Feng and Hou 
also extended their approach to mixtures of active (thermal AOUP) and passive particles \cite{Feng2018}.
They found that such mixtures can give rise to reentrant behavior upon increasing the active 
component fraction, provided that the activity is sufficiently weak and that the 
active and passive species have different particle sizes. Note that this reentrance effect is 
distinct from the earlier MCT predictions of Ding \textit{et al}.\ \cite{Ding2015}.

The 2017 work by Liluashvili, Onody, and Voigtmann constitutes the first
active-MCT work in which the translational and rotational degrees of freedom
are explicitly coupled and treated on an equal footing \cite{Liluashvili2017}. They focused on
two-dimensional ABPs and performed a detailed analysis of the theoretical
predictions for active hard particles. It was found that increasing the
self-propulsion speed generally causes the material to become more liquid-like;
this predicted active fluidization effect grows monotonically with increasing
activity or inverse rotational diffusion constant, and shifts the
glass-transition density to higher values. Interestingly, they also presented a
non-trivial glass-transition phase diagram in the three-dimensional parameter
space of density, self-propulsion speed, and rotational diffusivity, which is
reminiscent of the phase diagram proposed by Bi \textit{et al}.\ for the ABP
Voronoi model \cite{Bi2016}. This thus establishes an intriguing link between the dynamics of
living confluent tissues and the first-principles-based statistical physics of
glass-forming active liquids.  Furthermore, Liluashvili \textit{et al}.\ 
discussed the difference between the \textit{monotonic} activity dependence of
their ABP work and the \textit{non-monotonic} behavior of athermal AOUPs
reported by Szamel; they attributed this difference to the absence of thermal
Brownian noise in AOUPs, suggesting that finite thermal diffusive motion
generally makes caging less effective.

Very recently, Szamel also presented a new MCT version for thermal ABPs which
is based on a similar derivation as his AOUP work, i.e.\ the rotational degrees
of freedom are effectively integrated out \cite{Szamel2019}. It was predicted that at short times
the self-propulsion always speeds up the relaxation; at long times the
relaxation depends explicitly on the steady-state structure factor and the
correlation function of steady-state currents. Although numerical results were
not presented, it was concluded that the theory should reduce to the standard
MCT for passive glass formation in the limit of vanishing self-propulsion. If
thermal fluctuations are neglected, the theory becomes equivalent to the
athermal AOUP theory \cite{Szamel2016}, while in the limit of high rotational diffusivity the
theory coincides with the passive MCT at a temperature equal to the effective
temperature. It was noted that the presence of non-trivial currents may wash
away the ergodicity-breaking transition in active glassy matter; future work
will hopefully shed more light on the precise role of activity within this
first-principles framework.

Another important line of MCT research concerns the study of glassy dynamics in
the limit of infinite spatial dimensions $d$. While this limit is rather
abstract from an experimental point of view, the case $d\rightarrow \infty$
holds great relevance in theoretical physics, owing to the fact that it
constitutes an exact mean-field limit.  Two independent 2010 studies
\cite{Ikeda2010,Schmid2010} have found that, perhaps disappointingly, standard
MCT does not become exact for $d\rightarrow \infty$. In a landmark paper from
2016, however, Zamponi and co-workers managed to solve the exact dynamics of
glassy hard spheres in infinite dimensions \cite{Maimbourg2016}. Interestingly,
their final equation is somewhat similar but not identical to MCT. A subsequent
2017 study by Szamel also presented an alternative, physically motivated
derivation of this result \cite{Szamel2017}.  Impressively, Agoritsas,
Maimbourg, and Zamponi recently developed a dynamical $d \rightarrow \infty$
mean-field theory \cite{Agoritsas2019a,Agoritsas2019} that also applies to
non-equilibrium scenarios, including glassy aging and active (Ornstein-Uhlenbeck) dynamics
\cite{Agoritsas2019a}. Although explicit results for active glasses were not
presented, future work will hopefully shed more light on the role of activity
in a well-defined mean-field limit.  Ultimately, approaches such as a perturbative
expansion in $1/d$ may possibly allow one to extend these results to
finite-dimensional and experimentally realizable active glassy systems.
The reader is also referred to Ref.\ \cite{Agoritsas2019a} if she is interested in a broader discussion 
on the relevance of high-dimensional studies in the context of glassy dynamics. 

We finally mention another class of non-equilibrium materials that is closely
related to active fluids, namely driven granular matter.  Between 2010 and
2013, Kranz, Sperl, and Zippelius developed an MCT for driven granulates by
modeling the activity as a driving amplitude that gives rise to random particle
kicks \cite{Kranz2010,Sperl2012,Kranz2013}. It was found that the critical glass transition systematically shifts to
higher densities as the dissipation due to inelastic particle collisions
increases. The properties within the glass phase were also qualitatively and
quantitatively affected by the degree of dissipation. As already noted in Sec.\
\ref{sec:exptsynth}, such driven materials may be studied experimentally by, e.g., placing
granular particles on an air-fluidized or vibrating table, and it is hoped that
experiments in this field will be realized in the near future.

\subsection{Random First Order Transition Theory and the Stokes-Einstein relation}

We now discuss two other recent theoretical approaches
to describe the glassy physics of active matter, including an active version of
RFOT and an analytic treatment of Stokes-Einstein violation in athermal AOUPs.
Let us first recall the standard RFOT scenario, which is a popular theoretical
framework of the glass transition that is rooted in spin-glass theory \cite{Kirkpatrick1987,Kirkpatrick1987a}. Briefly,
RFOT amounts to a combination of the \textit{dynamical} scenario of MCT at
relatively high temperatures and \textit{thermodynamic} arguments at lower
temperatures.  In particular, the low-temperature system is described as a
mosaic of (Adam-Gibbs-like) amorphous domains with a characteristic
configurational entropy. The sizes of these domains are determined by their
stability against thermally activated fluctuations and grow progressively upon
deeper supercooling. The domains can transform among themselves by overcoming
their interfacial free energy due to thermal activation, and the time scale for
such rearrangements is defined as the structural relaxation time $\tau$. We
refer the reader to Refs.\ \cite{Lubchenko2007, Kirkpatrick2014, Biroli2009, Berthier2011a, Langer2014} 
for a more detailed description of the theory, including its successes
and limitations.

In 2018, Nandi \textit{et al}.\ \cite{Nandi2018} extended the RFOT framework to non-equilibrium
active systems in a phenomenological manner.  Specifically, they incorporated
the effect of activity in the expression for the configurational entropy; this
effect, in turn, was assumed to arise from an activity-induced change in the
(mean-field) potential energy, which should be valid only for small deviations
from equilibrium.  They studied the RFOT dynamics for two different flavors of
the active Ornstein-Uhlenbeck process, and concluded that a larger
self-propulsion strength always inhibits glassiness. The role of the
persistence time, however, was found to be more subtle and dependent on the
microscopic details of the activity. Notably, despite the simplifying
approximations in their theory, Nandi \textit{et al}.\ could account for the
seemingly paradoxical simulation results of Flenner \textit{et al}.\ \cite{Flenner2016} and
Mandal \textit{et al}.\ \cite{Mandal2016} regarding the effect of activity on the fragility.
Indeed, for the active-particle model used by Mandal \textit{et al}.\, the
material becomes stronger as the activity increases, while the athermal AOUP
model of Flenner and co-workers yields more fragile behavior as the
persistence time grows.  The correct prediction of both phenomena (using two
different active-RFOT models) highlights the fact that microscopic and
material-specific details of the self-propulsion mechanism must be carefully
taken into account when addressing glassy dynamics in active systems. 

Let us finally return to the topic of Stokes-Einstein violation--an important
feature that is observed in most glass-forming materials.  Although at a
phenomenological level the decoupling of the diffusion constant and the
viscosity (or relaxation time) may appear to be similar in passive and active
glassy materials--as reported by e.g.\ Flenner \textit{et al}.\ \cite{Flenner2016}--the underlying physics may be fundamentally different.
Indeed, in passive equilibrium systems it is rather surprising that the
Stokes-Einstein relation breaks down at low temperatures; in
non-equilibrium active matter, however, there is \textit{a priori} no reason to
assume that the Stokes-Einstein relation should hold at any temperature.  The latter
argument stems from the fact that the injection and dissipation of energy in
active matter are uncorrelated: injection of energy arises from the conversion
of some form of stored energy (either supplied internally or externally), while
dissipation is due to friction with the solvent. This is to be contrasted with
passive Brownian motion, in which both the movement and dissipative friction of
colloidal particles are governed by the same surrounding medium.

In 2016, Fodor \textit{et al}.\ theoretically studied this aspect in detail for
a system of athermal AOUPs \cite{Fodor2016}. They found that for sufficiently small persistence
times, the active system behaves as an effective equilibrium system in the
sense that detailed balance, time-reversal symmetry, and an effective
fluctuation-dissipation theorem hold. As the persistence time increases,
however, the material is driven further away from equilibrium, resulting in the
breaking of detailed balance and the lack of a generalized Stokes-Einstein
relation between damping and fluctuations. Thus, Stokes-Einstein violation
should always occur in strongly active (OUP) systems.  Although the work of
Fodor \textit{et al}.\ was not specifically motivated by glassy physics, their
work does carry implications for the study of glass formation. In particular,
it will be highly interesting to explore whether Stokes-Einstein violation in
active glassy matter is also linked to the emergence of dynamical
heterogeneity--as is generally believed to be the case for passive
glass-forming matter--, and to what extent the microscopic origins of
Stokes-Einstein violation in passive glass formers are fundamentally distinct
from non-equilibrium active materials.

\section{Conclusions and outlook}
\label{sec:concl}

This review has sought to provide an overview of the rapidly growing body of
literature on active glassy matter, focusing on the manifestation of the main
hallmarks of glassy dynamics in active systems from an experimental, numerical
simulation, and theoretical perspective, respectively. Although the field has
emerged only in recent years, it is already unambiguously clear that dense
active matter--either of a synthetic or biological nature--shares many
similarities with the phenomenology of conventional glass-forming materials. We
briefly recapitulate the key evidence regarding the five aspects of glassy
physics in non-equilibrium active matter below.

\begin{enumerate}[label=(\roman*)]
\item \textit{Dynamical slowdown.} As in equilibrium supercooled liquids,
disordered active matter can undergo kinetic arrest at sufficiently high
densities. Indeed, this is universally observed in both synthetic model systems
and in living cells; for the latter, a dramatic dynamical slowdown has been
reported at both the intracellular and intercellular level in various cell
types. Depending on the details of the self-propulsion mechanism, however,
activity can both promote and inhibit the slowdown.  In all cases, and as in
conventional glass formers, the orders-of-magnitude growth of the viscosity or
relaxation time is accompanied by only subtle changes in the material's
microstructure.  For synthetic active model glass formers, small structural
changes appear in the radial distribution or static structure factor which are
unique to the non-equilibrium material, i.e., they generally cannot be mapped
onto the structure of an effective equilibrium system.  For living confluent
cell layers, the geometric cell shape appears to be an important and unique
structural signature of glassiness. Curiously, there is compelling evidence
that living cells also employ activity as a means to control their own
relaxation time, e.g.\ by switching from a fluid-like to glass-like dormant
state under unfavorable environmental conditions. The apparent lack or delay of
a dynamical slowdown in confluent cells has also been associated with
pathological conditions such as asthma.

\item \textit{Fragility.} Both synthetic and living active materials can
exhibit varying degrees of fragility, ranging from strong Arrhenius-type to
fragile super-Arrhenius-type behavior. Depending on the microscopic details of
the system, non-equilibrium activity can both increase and decrease the
fragility. In living cells, the intracellular cytoplasm appears to switch from
fragile to strong as the metabolic activity level increases; active confluent
cell layers tend to behave as fragile materials.  The biological relevance of
fragility in living cells and tissues, as well as the experimental realization
of activity-controlled fragility in synthetic materials, still remains to be
established.

\item \textit{Dynamical heterogeneity.} As in conventional glass formers, both
synthetic and living active materials exhibit growing dynamically heterogeneous
behavior upon approaching the glass transition. Thus far, the role of activity
appears to amount to a similar phenomenology as in passive equilibrium systems:
as the relaxation time increases--either by activity or supercooling--, the
dynamics also becomes increasingly more heterogeneous.  In a biological
context, it has been hypothesized that the emergence of dynamically
heterogeneous domains within a primary tumor may partially underlie the
formation of metastatic cell clusters. 

\item \textit{Stokes-Einstein violation.} The manifestation of Stokes-Einstein
violation in active glassy materials has not yet been studied in great detail,
but the first results on synthetic model systems suggest that the behavior is
rather similar to conventional glass formers.  More
specifically, at high effective temperatures and low persistence times, the
Stokes-Einstein relation appears to hold. By departing from equilibrium through
the increase in persistence time, the diffusion constant and relaxation time
become more decoupled. Although the microscopic origin of Stokes-Einstein
violation in passive supercooled liquids is still debated, and may possibly be
associated with the emergence of dynamical heterogeneity, in active matter the
origin stems from the inherent lack of a fluctuation-dissipation theorem far
away from equilibrium. The latter presumably can occur even in
the absence of dynamical heterogeneity.  It remains unclear whether
Stokes-Einstein violation in active and passive glassy matter may ultimately be
captured in a single unifying framework, to what extent dynamical heterogeneity
plays a universal underlying role, and to what extent living glassy cells will
conform to the same physics as synthetic model systems.

\item \textit{Aging.} Owing to the non-Hamiltonian nature of active matter, the
conventional interpretation of physical aging as a gradual approach to a deeper
energy minimum is \textit{a priori} inapplicable to active glasses. Thus far,
only one simulation study has reported aging in athermal active glassy matter;
in this work, the observed aging behavior was interpreted as an activity-driven
approach to a more \textit{mechanically stable} state that effectively acts as
an absorbing state. It remains to be established whether,
e.g., the presence of thermal fluctuations in active matter will yield a
similar phenomenology as aging in passive thermal glasses. In experiments on
confluent cell layers, aging was found to be manifested in various physical
properties, including the gradual densification of the cell sheet, and the
maturation of cell-cell and cell-substrate contacts within the layer.  More
work is needed to clarify the main fundamental differences in physical aging
between active and non-active glassy matter, and between synthetic and living
active glasses.  
\end{enumerate}
 
We end by discussing several interesting open questions and possible directions
for future research. Regarding the most striking feature of glass formation,
i.e.\ the spectacular slowdown of the dynamics, it remains to be established
which microstructural material properties ultimately underlie this behavior.
Although the structure-dynamics link has still not been rigorously elucidated
for conventional supercooled liquids, it is plausible that this link will be
even more complex for non-equilibrium active glassy materials. Indeed, activity
is known to affect both the structural and dynamical properties in a
non-trivial manner, and the causal relation between these two effects is still
unknown. It will therefore be highly interesting to explore whether the
presence of self-propulsion forces primarily influences only the structure,
only the dynamics, or both. Furthermore, although a glassy slowdown has been
widely observed in both living and non-living active systems, the respective
microstructures of these materials have thus far been quantified in
fundamentally different ways. Future work should clarify whether
liquid-state-theory-based concepts, such as the radial distribution function,
can also be successfully applied to quantify relevant structural changes in,
e.g., confluent cell layers. If so, a theoretical connection may be established
between the (first-principles-based) statistical physics of active glassy
liquids and biological glassy phenomena. This raises the intriguing prospect
that, ultimately, the elusive structure-dynamics link in living and non-living
active matter can be captured in a single theoretical framework.

Fragility, i.e.\ the abruptness with which vitrification takes place, has
historically provided an important and unifying concept to classify seemingly
disparate classes of (passive) glass-forming systems. Indeed, it is widely
believed that a rigorous understanding of its microstructural origins will be
paramount to achieve a universal description of glass formation. Since
non-equilibrium activity has now been shown to affect the fragility in a
non-trivial manner--both in synthetic and in living active systems--, the
natural question arises whether the fundamental physical properties underlying
a material's fragility will be similar or distinct in active and passive
glasses. As in the case of the dynamical slowdown, this would also carry
implications for the development of a general and quantitative framework that
can rationally relate structural to dynamical properties, both for living and
non-living active systems.  
 
From a more applied point of view, fragility is also known to impact both the
processability and functionality of a material; for example, strong glass
formers are generally easier to (re-)shape and recycle, while fragile glass
formers can be used as, e.g., ultrafast phase-changing materials. We here
propose that the complex role of activity on a material's fragility may also
hold relevance in applied materials science and biology. Specifically, we pose
the following two questions: i) can the addition of self-propulsion forces be
exploited to create novel synthetic active materials with an \textit{externally
tunable} fluidization/solidification response, and ii) do living cells, both at
the intra- and intercellular level, invoke activity as a means to
\textit{control} their own fragility? Regarding the latter, it might be
possible that the up- or down-regulation of metabolic activity allows cells to
control the transport properties within their cytoplasm, and to regulate the
abruptness of solidification in collective cell migration processes such as
wound healing and embryonic development.  More generally, it will be highly
interesting to explore if and how a departure from equilibrium will enable
novel dynamic control mechanisms in functional glassy matter.

Our current understanding of dynamical heterogeneity, i.e.\ the increasingly
cooperative nature of particle rearrangements near the glass transition, is
still far from complete; however, it is broadly accepted that the emergence of
dynamically heterogeneous behavior is concomitant to the slowdown in structural
relaxation. As already discussed above, the establishment of a framework that
can accurately predict the full glassy dynamics, including the degree of
dynamical heterogeneity, for a given material composition and structure will
constitute a major breakthrough in the field of condensed matter physics. For
active materials, it is not yet clear whether the presence of non-equilibrium
activity will generally add a fundamentally new ingredient to the physics of
dynamical heterogeneity; indeed, the first studies suggest that the phenomenology
in passive and active glass-forming materials is rather similar. Nonetheless,
it is plausible that active forces, and especially actively aligning
interactions, can induce non-trivial spatiotemporal correlations and
coordinated movements among particles that would be absent in the passive
counterpart.  Future work is needed to establish how the manifestation of
dynamical heterogeneity, and the possible link with the underlying
microstructure, is dependent on the details of the self-propulsion mechanism.
As in the case of fragility, it will be interesting to study whether living
cells also exploit activity--either directly in the dynamics or indirectly via
changes in the microstructure--to effectively control the size and growth of
dynamically heterogeneous regions. A speculative but intriguing possibility is
that the emergence of such transiently mobile domains within cancerous cell
tumors partially contributes to the formation of cell clusters with enhanced
mobility, effective initializing the onset of metastasis. 

Curiously, the breakdown of the Stokes-Einstein relation, i.e.\ the decoupling
of the temperature-dependent diffusion constant and relaxation time, is
currently better understood for active materials than for passive glass-forming
materials. The presence of strong non-equilibrium forces gives rise to the
breaking of detailed balance and the lack of a fluctuation-dissipation theorem,
thus rendering the Stokes-Einstein relation generally inapplicable to active
matter. Conversely, for passive supercooled liquids, it has been hypothesized
that Stokes-Einstein violation is essentially a manifestation of dynamical
heterogeneity; it remains to be established whether an increased violation of
the Stokes-Einstein relation in active matter is also generally accompanied by
increasingly dynamically heterogeneous behavior, and whether living active
cells will also conform to this phenomenology. Even if at a phenomenological
level the decoupling of the diffusion constant and relaxation time will be
manifestly equivalent, the physical origins of this decoupling need not be the
same in equilibrium and far-from-equilibrium systems. Future work will
hopefully shed more light on these origins, in particular on the possible links
with the underlying microstructure and the degree of dynamical heterogeneity.

Arguably the least studied glassy property in active matter is aging, i.e.\ the
explicit dependence of a material on its age. Although proof-of-principle for
this non-equilibrium phenomenon has recently been established for active
matter, there is still a multitude of interesting open questions that remain to
be answered. For example, can non-equilibrium activity act as an effective
proxy for thermal fluctuations in passive aging glasses? Can the notion of a
mechanical energy landscape and an absorbing-state formalism be generalized to
other systems?  How does the applied aging protocol affect the active-aging
behavior? Can specific features of physical aging in passive glass formers,
such as a logarithmic decay and a time-versus-waiting-time superposition
principle, also be observed in active glasses? And to what extent can the aging
behavior of synthetic active glasses be translated to the aging of living
cells? 

Finally, we note another disparate field of research in which the study of
active glassy dynamics and aging may potentially find applicability, namely
neural networks. For example, in so-called Hopfield networks \cite{Hopfield1982}, a memory pattern 
is encoded as a local energy minimum that acts as an attractor in a rugged energy landscape.
Passive spin glasses form a suitable realization of such networks \cite{Amit1985}, but it might
be possible to extend the scope and functionality of Hopfield networks by
adding activity as a new degree of freedom. The question how efficiently
different aging protocols drive active (spin) glasses into a basin of
attraction may thus also offer new possibilities for the design and
implementation of 'active' neural networks.

\acknowledgments
It is a pleasure to thank Ludovic Berthier, Grzegorz Szamel, Roberto Cerbino, and Atsushi Ikeda
for interesting discussions. Vincent Debets is gratefully acknowledged for his 
help in preparing Figure 3.

%

\end{document}